\documentclass[11pt]{article}

\usepackage{mathrsfs}
\usepackage{amssymb}
\usepackage{chet}

\newcommand{\lsp}{\hspace{1pt}}
\newcommand{\dd}{\ensuremath{d^{\lsp d}}}
\newcommand{\dtwo}{\ensuremath{d^{\hspace{0.5pt}2}}}
\newcommand{\dfour}{\ensuremath{d^{\hspace{0.5pt}4}\hspace{0.5pt}}}

\newcommand{\ddp}{\ensuremath{d^{\lsp d+1}}}
\newcommand{\DeltaW}{\Delta^{\hspace{-2pt}\text{W}}}
\newcommand{\Deltam}{\Delta^{\hspace{-1pt}m}}
\newcommand{\DeltaWhat}{\hat{\Delta}^{\hspace{-2pt}\text{W}}}
\newcommand{\Lag}{\mathscr{L}}
\newcommand{\Lagloc}{\Lag_{\text{loc}}}
\newcommand{\Sloc}{S_{\text{loc}}}
\newcommand{\Ham}{\mathscr{H}}
\newcommand{\CHam}{\mathcal{H}}
\newcommand{\CHamhat}{\widehat{\mathcal{H}}}
\newcommand{\CP}{\mathcal{P}}
\newcommand{\CO}{\mathcal{O}}
\newcommand{\CPhat}{\widehat{\mathcal{P}}}
\newcommand{\Shat}{\widehat{S}}
\newcommand{\nus}{\nu\hspace{-0.7pt}}
\newcommand{\gm}{{\tilde{g}}}
\newcommand{\hm}{{\tilde{h}}}
\newcommand{\Lien}{\pounds_{\hspace{-1pt} n}\hspace{0.3pt}}
\newcommand{\Liea}{\pounds_{\hspace{-1pt} a}\hspace{0.3pt}}
\newcommand{\Liet}{\pounds_{\hspace{-1pt} t}\hspace{0.3pt}}
\newcommand{\LieN}{\pounds_{\hspace{-1pt} N}\hspace{0.3pt}}
\newcommand{\Liebeta}{\pounds_{\hspace{-1pt} \beta}\hspace{0.3pt}}
\newcommand{\hmdot}{{\dot{\hm}}}
\newcommand{\Kdot}{{\dot{K}}}
\newcommand{\phidot}{{\dot{\phi}}}
\newcommand{\Sigmadot}{{\dot{\Sigma}}}

\preprint{MIT-CTP-4703}

\date{August 2015}

\title{Holographic Trace Anomaly and\\\vspace{3pt} Local Renormalization
Group}

\author{Srivatsan Rajagopal,$^{\!a}$ Andreas Stergiou$^{b}$ and Yechao
Zhu$^{c}$ \emails{srivat91@mit.edu, andreas.stergiou@yale.edu,
eltonzhu@mit.edu}}

\affiliation{$^a$Center for Theoretical Physics, Massachusetts Institute
of Technology,\\\vspace{-3pt}Cambridge, Massachusetts 02139, USA\\
$^b$Department of Physics, Yale University, New Haven, CT 06520, USA\\
$^c$Department of Physics, Massachusetts Institute
of Technology,\\\vspace{-3pt}Cambridge, Massachusetts 02139, USA}

\abstract{The Hamilton--Jacobi method in holography has produced important
results both at a renormalization group (RG) fixed point and away from it.
In this paper we use the Hamilton--Jacobi method to compute the holographic
trace anomaly for four- and six-dimensional boundary conformal field
theories (CFTs), assuming higher-derivative gravity and interactions of
scalar fields in the bulk. The scalar field contributions to the anomaly
appear in CFTs with exactly marginal operators. Moving away from the fixed
point, we show that the Hamilton--Jacobi formalism provides a deep
connection between the holographic and the local RG.  We derive the local
RG equation holographically, and verify explicitly that it satisfies Weyl
consistency conditions stemming from the commutativity of Weyl scalings. We
also consider massive scalar fields in the bulk corresponding to boundary
relevant operators, and comment on their effects to the local RG equation.}

\begin{document}

\maketitle

\toc

\newsec{Introduction}
In a conformal field theory (CFT) defined in flat space the trace of the
stress-energy tensor vanishes. Despite this, in an even-dimensional CFT
considered in curved space there is an anomaly and the stress-energy tensor
is no longer traceless due to curvature contributions~\cite{Capper:1974ic}.
This trace anomaly has been extensively studied in field theory and beyond
(see~\cite{Duff:1993wm} for a nice review), and, soon after the discovery
of the AdS/CFT correspondence \cite{Maldacena:1997re, Gubser:1998bc,
Witten:1998qj}, it was also considered in the context of holography. In
particular, Henningson and Skenderis (HS) provided a holographic derivation
of the trace anomaly of the boundary two- four- and six-dimensional
CFT~\cite{Henningson:1998gx} by studying the divergences of the
supergravity action close to the AdS boundary. The four-dimensional result
was also obtained later by de Boer, Verlinde and Verlinde (dBVV) using the
Hamiltonian formulation of gravity and Hamilton--Jacobi
theory~\cite{deBoer:1999xf, deBoer:2000cz}, while the $d=6$ result was
obtained using the dBVV method in~\cite{Fukuma:2000bz}. Additionally, the
dBVV method was used in~\cite{Nojiri:2000mk} to compute the $d=8$
holographic anomaly. The Hamilton--Jacobi method was also considered by
Papadimitriou and Skenderis~\cite{Papadimitriou:2004ap,
Papadimitriou:2004rz}.

The results of HS and dBVV were obtained with Einstein gravity in the bulk,
but this was extended to higher-derivative gravity in~\cite{Nojiri:1999mh,
Blau:1999vz} and \cite{Fukuma:2001uf}. A scalar field $\phi$ has also been
considered in the bulk with its kinetic term, while more general situations
involving also an axion have been examined in~\cite{Papadimitriou:2011qb,
Jahnke:2014vwa}. The flat space limit of the corresponding contributions to
the anomaly has been considered in~\cite{deHaro:2000xn} in any even
dimension, while the full curvature-dependent contributions have been
computed in four dimensions in~\cite{Fukuma:2000bz}. The $\phi$-dependent
terms for massless $\phi$ correspond to contributions in CFTs with
conformal manifolds, and in $d=4$ they take the form of the Paneitz
operator in $d=4$\cite{Paneitz}, an operator first discussed by Fradkin and
Tseytlin~\cite{Fradkin:1981jc, Fradkin:1982xc, Fradkin:1983tg} and also
considered by Riegert~\cite{Riegert:1984kt}. The form of these
contributions is dictated by conformal covariance~\cite{Jack:2013sha}. For
general even dimension the conformal covariance properties of the
$\phi$-dependent part of the holographic trace anomaly were pointed out
in~\cite{Papadimitriou:2004ap, Papadimitriou:2004rz}. The case of massive
$\phi$ has been considered in~\cite{Bianchi:2001kw}.

In this work we extend these results in $d=4,6$ by considering
higher-derivative gravity in the bulk, with higher-derivative quadratic
interactions of a scalar field. Our bulk action is given in
\eqref{BulkAction} below, and we also consider the required boundary action
\eqref{BoundaryAction}. Following the method of \cite{Fukuma:2001uf} we
obtain our results for the trace anomaly in equations \eqref{FourdTA} and
\eqref{SixdTA} in $d=4,6$ respectively. In $d=4$ the new higher-derivative
interactions of the scalar contribute to the trace anomaly in accord with
the Paneitz operator~\cite{Paneitz, Fradkin:1981jc, Fradkin:1982xc,
Fradkin:1983tg, Riegert:1984kt}, just like the standard kinetic term in the
bulk. In $d=6$ the higher-derivative gravity terms and the standard kinetic
term of $\phi$ have not been considered before. For the $\phi$-dependent
contributions we find that the kinetic term gives rise to the Branson
operator~\cite{Branson}, a conformally covariant operator defined in $d=6$,
while the higher-derivative terms contribute to the Branson operator, but
also give rise to two more conformally covariant operators quadratic in
$\phi$ and involving the Weyl tensor. These operators were shown to appear
in CFTs with marginal operators in $d=6$ in~\cite{Osborn:2015rna}.

The trace anomaly in $d=6$ CFTs in curved space contains three Weyl
invariant contributions, with coefficients $c_1, c_2,c_3$, as well as the
Euler term, with coefficient $a$. The parameter $c_3$ appears in the
two-point function of the stress-energy tensor two-point function in flat
space, while $c_1, c_2$ show up in the three-point function.  Using
positivity of energy flux in lightlike directions it was shown
in~\cite{Hofman:2008ar} that one can obtain bounds on the parameters
appearing in the three-point function of the stress-energy tensor. These
bounds were understood holographically in~\cite{Hofman:2009ug}, where they
were shown to arise by causality considerations in the bulk. For the $d=6$
case these bounds were considered in~\cite{deBoer:2009pn, Camanho:2009vw,
Buchel:2009sk} for Gauss--Bonnet gravity in the bulk. In this paper we
extend the result of~\cite{deBoer:2009pn, Camanho:2009vw, Buchel:2009sk}
for general higher-derivative gravity in the bulk using our result
\eqref{SixdTA}.

Outside a conformal fixed point, Osborn has introduced a systematic
treatment of the trace anomaly incorporating efficiently renormalization
effects of composite operators~\cite{Osborn:1991gm}. In Osborn's analysis a
background metric $\gamma_{\mu\nu}$ is introduced and the couplings $g^I$
are promoted to spacetime-dependent sources for the corresponding composite
operators $\CO_I$.  Besides curvature-dependent counterterms required for
finiteness, one needs to also consider counterterms containing derivatives
on $g^I$~\cite{Jack:1990eb, Grinstein:2015ina}.  Then, a local
renormalization group (RG) equation can be derived, valid along the RG
flow. This corresponds to a local version of the Callan--Symanzik equation,
and yields an expression of the form
\eqn{T^\mu{\!}_{\mu}=\beta^I\CO_I+(\text{terms with
derivatives on }\gamma_{\mu\nu}, g^I)}[OsbTA]
for the trace of the stress-energy tensor $T_{\mu\nu}$. The terms with
derivatives on $\gamma_{\mu\nu}$ and $g^I$ in \OsbTA contain coefficients
which may be related to flat-space correlation functions involving
$T_{\mu\nu}$ and $\CO_I$.

Osborn further considered \OsbTA and the response of the field theory under
Weyl scalings, and derived consistency conditions in $d=2,4$ stemming from
the Abelian nature of the Weyl group. These are similar to the well-known
Wess--Zumino consistency conditions~\cite{Wess:1971yu}, and have been a
subject of interest recently in $d=4$~\cite{Jack:2013sha, Baume:2014rla},
as well as in $d=6$ and more generally in any even
$d$~\cite{Grinstein:2013cka}. The main driving force has been a consistency
condition of the form
\eqn{\mu\lsp\frac{d\tilde{a}}{d\mu}=G_{IJ}\lsp\beta^I\hspace{-1pt}
\beta^J\,,}[atcc]
where $\mu$ is the RG scale, found by Osborn in $d=2,4$ and shown
in~\cite{Grinstein:2013cka} to appear in any even $d$. Equation \atcc ties
the monotonicity of the RG flow of a quantity $\tilde{a}$, related to the
coefficient of the Euler term in the trace anomaly, to the sign of a
symmetric tensor $G_{IJ}$. In $d=2$ a positive-definite $G_{IJ}$ was
found by Osborn~\cite{Osborn:1991gm}, thus rederiving Zamolodchikov's
$c$-theorem~\cite{Zamolodchikov:1986gt}, while in $d=4$ only a perturbative
analog was obtained~\cite{Jack:1990eb}. In $d=6$ the sign of $G_{IJ}$
was found to be negative in multiflavor $\phi^3$
theory~\cite{Grinstein:2014xba, Grinstein:2015ina}.

Despite their obvious interest from the field theoretic point of view,
Osborn's local RG and consistency conditions have received limited
attention from the holographic side. Erdmenger developed the subject to
some extent in~\cite{Erdmenger:2001ja}, but the results derived there do
not illustrate the deep connection of Osborn's formalism with the dBVV
formulation of the holographic RG. In this paper we show that the flow
equation of dBVV contains Osborn's local RG equation. In $d=4$ and with
Einstein gravity and a massless scalar field in the bulk we compute
holographically quantities in the local RG equation \OsbTA like $\tilde{a}$
and $G_{IJ}$ mentioned above. These quantities are related to the local
divergent part of the supergravity action close to the boundary.
Furthermore, we verify that all Weyl consistency conditions derived by
Osborn in $d=4$ are satisfied by the holographic result. We also consider
bulk massive scalar fields, and comment on their contributions the anomaly.

This paper is organized as follows. In the next section we describe for
completeness the formalism of dBVV in the higher-derivative case. We derive
all necessary results needed for the computation of the holographic trace
anomaly in section~\ref{secTA}. In section~\ref{secOsb} we illustrate the
relation of the holographic RG to the local RG, and derive an expression
for the holographic trace anomaly away from the fixed point. In this
section we also comment on the $a$-theorem-like consistency condition~\atcc
in $d=4,6$, and discuss the effects on the anomaly originating from massive
scalar fields in the bulk. We also include appendices on details of the ADM
decomposition, the boundary terms, definitions of curvature tensors in
$d=6$, and results in $d=4$ for the coefficients of the anomaly terms away
from the fixed point assuming Einstein gravity and a kinetic term for the
massless scalar field $\phi$ in the bulk.

\newsec{Higher-derivative dilaton gravity}
We consider classical dilatonic gravity on an asymptotically-AdS manifold
$M_{d+1}$ with metric $\gm_{\mu\nu}$. The bulk action is taken to be
\eqn{S_B=\int_{M_{d+1}}\ddp x\sqrt{\gm}\,\big(\Lag_B^\gm+\Lag_B^\phi\big)\,
,}[BulkAction]
where $\gm$ is the determinant of $\gm_{\mu\nu}$ and
\eqna{\Lag_B^\gm&=2\lsp\Lambda-R-a\lsp R^2-b\lsp R^{\mu\nu}R_{\mu\nu}
-c\lsp R^{\mu\nus\rho\sigma}R_{\mu\nus\rho\sigma}\,,\\
\Lag_B^\phi&=\tfrac12\lsp\partial^\mu\phi\lsp\partial_\mu\phi
+e\lsp R\lsp\lsp\partial^\mu\phi\lsp\partial_\mu\phi
+f R^{\mu\nu}\lsp\partial_\mu\phi\lsp\partial_\nu\phi
+g\lsp\nabla^2\phi\lsp\nabla^2\phi
+h\lsp\nabla^\mu\partial^\nu\phi\lsp\nabla_\mu\partial_\nu\phi\,.
}[BulkTerms]
Here we allow terms quadratic in $\phi$ with up to two derivatives on
$\phi$. The manifold $M$ has a $d$-dimensional boundary $\partial M$ and we
also have the boundary action
\eqn{S_{\partial}=\int_{(\partial M)_d}\dd y \sqrt{\hm}\,
\big(\Lag_{\partial}^\hm + \Lag_{\partial}^\phi\big)\,,}[BoundaryAction]
where $\hm_{ij}$ is the induced metric and
\eqna{\Lag_\partial^\hm&=2\lsp K+x_1\lsp RK+x_2\lsp R^{ij}K_{ij}+x_3\lsp
K^3+x_4\lsp KK^{ij}K_{ij}+x_5\lsp K^{i}{\!}_{j}K^{j}{\!}_k K^k{\!}_{i}\,,\\
\Lag_\partial^\phi&=y_1\lsp K\lsp\partial^i\phi\lsp\partial_i\phi
+y_2\lsp K^{ij}\lsp\partial_i\phi\lsp \partial_j\phi
+y_3\lsp \Lien\phi\lsp\nabla^2\phi\,,}[BoundaryTerms]
where $K_{ij}$ is the extrinsic curvature, $K=\hm^{ij}K_{ij}$, and $\Lien$
is the Lie derivative along the vector $n^\mu$ normal to the boundary. The
first term in $\Lag_\partial^\phi$ is the Gibbons--Hawking--York term for
Einstein gravity \cite{Gibbons:1976ue, York:1972sj}. More comments on the
boundary terms can be found in Appendix \ref{appBound}. The case with
$e=f=g=h=y_1=y_2=y_3=0$ has been considered in \cite{Fukuma:2001uf,
Fukuma:2002sb}.

It is straightforward to work out the ADM form \cite{Arnowitt:1962hi} of
the action
\eqn{S=S_B-S_\partial\,.}[]
Using technology summarized in Appendix \ref{appADM} we can determine
\eqn{S=\int dr\int\dd y\sqrt{\hm}\, \Lag\,,\qquad
\Lag=\Lag_0^\hm+\Lag_0^\phi+\Lag_1^\hm+\Lag_1^\phi\,,}[HDAction]
where the radial coordinate $r$ is identified with the RG parameter of the
boundary theory and
\eqn{\frac{1}{N}\Lag_0^\hm=2\lsp \Lambda-R-K^2+K^{ij}K_{ij}\,,\qquad
\frac{1}{N}\Lag_0^\phi=\tfrac12\lsp\big(\partial^i\phi\lsp\partial_i\phi
+(\Lien\phi)^2\big)\,,}[Lag1]
\eqna{\frac{1}{N}\Lag_1^\hm&=-a\lsp R^2-b\lsp R^{ij}R_{ij}-c\lsp
R^{ijkl}R_{ijkl}\\
&\quad +\big((2\lsp a-x_1)K^2-2\lsp(3\lsp a-x_1)K^{ij}K_{ij}\big)R\\
&\quad+\big((2\lsp b+2\lsp x_1-x_2)KK^{ij}
-2\lsp(2\lsp b+4\lsp c-x_2)K^i{\!}_kK^{kj}\big)R_{ij}
+2\lsp(6\lsp c+x_2)K^{ik}K^{jl}R_{ijkl}\\
&\quad-(a+x_3)K^4+(6\lsp a-b+6\lsp x_3-x_4)K^2K^{ij}K_{ij} -(9\lsp
a+b+2\lsp c-2\lsp x_4)(K^{ij}K_{ij})^2\\
&\quad+(4\lsp b+4\lsp x_4-x_5)KK^i{\!}_jK^j{\!}_kK^k{\!}_i
-2\lsp(2\lsp b+c-3\lsp x_5)K^i{\!}_jK^j{\!}_kK^k{\!}_lK^l{\!}_i\\
&\quad +2\lsp(b+x_1)K\nabla^2 K+(8\lsp c+x_2)K^{ij}\nabla^2K_{ij}\\
&\quad -(4\lsp b+2\lsp x_1-x_2)K^{ij}\nabla_i\partial_jK+2\lsp(b-4\lsp
c-x_2)K^{ij}\nabla_j\nabla_kK^k{\!}_i\\
&\quad-\big((4\lsp a+b)\hm^{ij}\hm^{kl}+(b+4\lsp c)\hm^{ik}\hm^{jl}\big)
L_{ij}L_{kl}\\
&\quad+\big((4\lsp a-x_1)R\lsp\hm^{ij}+(2\lsp b-x_2)R^{ij}\big)L_{ij}\\
&\quad-\big((4\lsp a+3\lsp x_3)K^2-(12\lsp a+2\lsp
b-x_4)K^{ij}K_{ij}\big)L\\
&\quad-\big(2\lsp(b+x_4)KK^{ij}
-(4\lsp b+8\lsp c-3\lsp x_5)K^i{\!}_kK^{kj}\big)L_{ij}\,,}[LagTwoh]
and
\eqna{\frac{1}{N}\Lag_1^\phi&= e\lsp R\lsp\partial^i\phi\lsp\partial_i\phi
+fR^{ij}\lsp\partial_i\phi\lsp\partial_j\phi
+g\lsp\nabla^2\phi\lsp\nabla^2\phi
+h\lsp\nabla^i\partial^j\phi\lsp\nabla_i\partial_j\phi\\
&\quad-\big((e+y_1)K^2 -(3\lsp e+2\lsp y_1)K^{ij}K_{ij}
+(2\lsp e+y_1)L\big)\partial^k\phi\lsp\partial_k\phi\\
&\quad-\big((f-2\lsp y_1+y_2)KK^{ij}
-2\lsp(f+h+2\lsp y_2)\lsp K^i{\!}_kK^{kj}
+(f+y_2)L^{ij}\big)\partial_i\phi\lsp\partial_j
\phi\\
&\quad+\big(e\lsp R-(e-g)\lsp K^2+(3\lsp e+f+h)K^{ij}K_{ij}
-(2\lsp e+f)L\big)(\Lien\phi)^2\\
&\quad+2\lsp g\lsp K\Lien\phi\lsp(\Lien\Lien\phi-\Liea\phi)
+(g+h)(\Lien\Lien\phi-\Liea\phi)^2\\
&\quad+2\lsp (f+g) K\lsp\nabla^2\phi\,\Lien\phi
-2\lsp(f-h) K^{ij}\lsp\nabla_i\partial_j\phi\,\Lien\phi\\
&\quad +(2\lsp g+y_3)\lsp\nabla^2\phi\lsp(\Lien\Lien\phi-\Liea\phi)
+(2\lsp h-y_3)\lsp\partial^i\!\Lien\phi\,\partial_i\Lien\phi\\
&\quad+(2\lsp f-2\lsp y_1-y_3)K\lsp\partial^i\phi\,\partial_i\Lien\phi
-2\lsp(f+2\lsp h+y_2-y_3)K^{ij}\lsp\partial_i\phi\,\partial_j\Lien\phi\,.
}[LagTwophi]
In \LagTwoh and \LagTwophi we neglect terms in the right-hand side that are
total derivatives.

For the higher-derivative Lagrangian $\Lag$ of \HDAction we consider the
canonical variables $g_{ij}$, $K_{ij}$, $\phi$, $\Sigma=\Lien\phi$,
$\pi_{ij}$, $P_{ij}$, $\pi_\phi$, and $P_\Sigma$, with the usual
definitions
\eqn{\pi_{ij}=\frac{\partial\Lag}{\partial \hmdot{\vphantom{\hm}}^{ij}}\,,
\qquad P_{ij}=\frac{\partial\Lag}{\partial \Kdot^{ij}}\,,\qquad
\pi_\phi=\frac{\partial\Lag}{\partial\phidot}\,,\qquad
P_{\lsp\Sigma}=\frac{\partial\Lag}{\partial\Sigmadot}\,.}[HamEqOne]
Since $L_{ij}$ is linear in $\Kdot_{ij}$ it is easy to compute
\eqna{P^{ij}&=-2\lsp\big((4\lsp a+b)\hm^{ij}\hm^{kl}+(b+4\lsp
c)\hm^{ik}\hm^{jl}\big)L_{kl}\\
&\quad+\big((4\lsp a-x_1)R -(4\lsp a+3\lsp x_3)K^2
+ (12\lsp a+2\lsp b-x_4)K^{kl}K_{kl} \big)\hm^{ij}\\
&\quad+(2\lsp b-x_2)R^{ij} -2\lsp(b+x_4)KK^{ij}
+ (4\lsp b+8\lsp c-3\lsp x_5)K^i{\!}_kK^{kj}\\
&\quad -\big((2\lsp e+f)\lsp\Sigma^2+(2\lsp e+y_1)\lsp
\partial^k\phi\lsp\partial_k\phi\big)\hm^{ij}
-(f+y_2)\lsp\partial^i\phi\lsp\partial^j\phi\,.}[eqPL]
Equation \eqPL is solved for $L_{ij}$ by
\eqn{L_{ij}=L^\prime_{ij}(P-P^\phi)\,,\qquad
P^{\phi}_{ij}=-\big((2\lsp e+f)\lsp\Sigma^2
+(2\lsp e+y_1)\lsp\partial^k\phi\lsp\partial_k\phi\big)\hm_{ij}
-(f+y_2)\lsp\partial_i\phi\lsp\partial_j\phi\,,}[eqLP]
where
\eqna{L^\prime_{ij}(P)&=-\frac{1}{2\lsp(b+4\lsp c)}\big(P_{ij}
-(2\lsp b-x_2)R_{ij} +2\lsp (b+x_4)KK_{ij}
-(4\lsp b+8\lsp c-3\lsp x_5)K_{ik}K^k{\!}_j\big)\\
&\hspace{-0.7cm}+\frac{1}{2\lsp(b+4\lsp c)(d\lsp(4\lsp a+b)+b+4\lsp c)}
\Big((4\lsp a+b)P -\lsp\big(2\lsp b^2+4\lsp a\lsp(b-4\lsp c)+(b+4\lsp c)
\lsp x_1 -(4\lsp a+b)\lsp x_2\big)R\\
&\hspace{4.8cm}+\big(2\lsp b^2+4\lsp a\lsp(b-4\lsp
c)-3\lsp(b+4\lsp c)\lsp x_3 +2\lsp(4\lsp a+b)\lsp x_4\big)K^2\\
&\hspace{4.8cm}-\big(2\lsp b^2+4\lsp a(b-4\lsp c) +(b+4\lsp c)\lsp x_4
-3\lsp(4\lsp a+b)\lsp x_5 \big)K^{kl}K_{kl}\Big)\hm_{ij}\,.}[eqLprimeP]
Also, since $\Lien\Sigma$ is linear in $\Sigmadot$ we find
\eqn{P_{\lsp\Sigma}=2\lsp(g+h)\lsp(\Lien\Sigma-\Liea\phi)
+ 2\lsp g\lsp(\nabla^2\phi+K\lsp\Sigma)\,,}[PSigma]
which allows us to express
\eqn{\Sigmadot=N\left(\frac{1}{2\lsp(g+h)}\big(P_{\lsp\Sigma}-2\lsp
g\lsp(\nabla^2\phi+K\lsp\Sigma)\big)+\Liea\phi\right)+\LieN\Sigma\,.
}[SigmaPSigma]

The action \HDAction can now be written in the first order form
\eqna{S&=\int dr\int\dd y\sqrt{\hm}\,\big(\pi^{ij}(\hmdot_{ij}
-2\lsp N K_{ij}-\nabla_iN_j-\nabla_jN_i)
+\pi_\phi(\phidot-N\Sigma-\LieN\phi)+\Lag\big)\\
&=\int dr\int\dd y\sqrt{\hm}\,\big(\pi^{ij}\hmdot_{ij}+\pi_\phi\lsp\phidot
+P^{ij}\Kdot_{ij}+P_{\lsp\Sigma}\lsp\Sigmadot-\Ham\big)\,,}[ActionHam]
with
\eqn{\Ham=\pi^{ij}(2\lsp NK_{ij}+\nabla_iN_j+\nabla_j N_i)
+\pi_\phi(N\Sigma+\LieN\phi)+P^{ij}\Kdot_{ij}+P_{\lsp\Sigma}\lsp\Sigmadot
-\Lag\,,}[HamOne]
which can be brought to the form
\eqn{\Ham=N\CHam(\gm,\phi,K;\pi,\pi_\phi,P-P^\phi)
+N^i\CP_i(\gm,\phi,K;\pi,\pi_\phi,P-P^\phi)\,,}[HamMom]
for appropriate $\CHam$ and $\CP$ that can be easily worked out. For this
one needs to use \eqref{Ldefn} and \eqLP.

Requiring that the variation of $S$ vanishes gives us Hamilton's equations
and a constraint at the boundary, which can be satisfied by either
Dirichlet or Neumann boundary conditions for the variables $\hm_{ij}$,
$\phi$, $K_{ij}$ and $\Sigma$. In order to impose Dirichlet boundary
conditions for $\hm_{ij}$ and $\phi$, and Neumann boundary conditions for
$K_{ij}$ and $\Sigma$, the action $S$ needs to be modified appropriately.
This can be done by means of a canonical transformation so that instead of
$S$ we use
\eqna{\Shat&=S-\int\ddp x\,\Liet\big(\hspace{-0.5pt}\sqrt{\hm}
\lsp(P^{ij}K_{ij}+P_{\lsp\Sigma}\lsp\Sigma)\big)\\
&=\int dr\int\dd y \sqrt{\hm}\,\big(\pi^{ij}\hmdot_{ij}
+\pi_\phi\lsp\phidot-K_{ij}\dot{P}^{ij}-\Sigma\lsp\dot{P}_{\lsp\Sigma}
-N\CHamhat-N^i\CPhat_i\big)\,,}[eqShat]
with
\eqna{\CHamhat&=\CHam+K(K^{ij}P_{ij}+\Sigma\lsp P_{\lsp\Sigma})\,,\\
\CPhat_i&=\CP_i-\nabla_i(K^{jk}P_{jk}+\Sigma\lsp P_{\lsp\Sigma})\,.}[]
Now we can impose Dirichlet boundary conditions for $\hm_{ij}$ and $\phi$,
and Neumann boundary conditions for $K_{ij}$ and $\Sigma$.  As we observe
$\Shat$ in \eqShat does not contain derivatives of $N$ or $N^i$, and so
these act as Lagrange multipliers enforcing the Hamiltonian and momentum
constraints
\eqn{\CHamhat=0\qquad\text{and}\qquad\CPhat_i=0\,.}[HamMomConstraints]

To proceed we need to obtain an action defined at the boundary. Let
$\bar{\hm}_{ij}$, $\bar{\phi}$, $\bar{P}_{ij}$ and $\bar{P}_{\lsp\Sigma}$
be the solutions of $\delta\Shat=0$ with the appropriate boundary
conditions. Using these solutions in $\Shat$ and defining
$\bar{\hm}_{ij}(y,r=r_0)\equiv \hm_{ij}(y)$ etc., gives us the classical
action $S_c\big[\hm(y),\phi(y),P(y)-P^\phi(y),P_{\lsp\Sigma}(y)\big]$, and
we have
\eqn{\CHamhat_c(\hm,\phi,K;\pi,\pi_\phi,P-P_\phi)=0\qquad
\text{and}\qquad\CPhat_c(\hm,\phi,K;\pi,\pi_\phi,P-P_\phi)=0\,.}[HamMomConstraintsTwo]
The Hamiltonian and momentum constraints \HamMomConstraintsTwo can be
recast as equations for the reduced classical action $S_r$ defined as
\eqn{S_r\big[\hm,\phi\big]=S_c\big[\hm,\phi,0,0
\big]\,,}[]
where the Neumann boundary conditions for $K_{ij}$ and $\Sigma$ have been
used in $S_c$. The conjugate momenta on the boundary (fixed $r=r_0$) are
given by
\eqn{\pi|^{ij}=-\frac{1}{\sqrt{\hm}}\frac{\delta S_r}{\delta\hm_{ij}}\,,
\qquad \pi|_\phi=-\frac{1}{\sqrt{\hm}}\frac{\delta
S_r}{\delta\phi}\,.}[Hameq]

The constraints \HamMomConstraintsTwo can be recast as constraints on $S_r$
by using $S$ of \HDAction.  Starting with $\Lag_0^\hm+\Lag_0^\phi$, which
we write in the form $\pi^{ij}\hmdot_{ij}+\pi_\phi\lsp\phidot -\Ham_0^\hm
-\Ham_0^\phi$, we can use \HamEqOne to determine
\eqn{\pi_{ij}=\lsp K_{ij}-K\hm_{ij}\,,\qquad
\pi_\phi=\Lien\phi\,,}[eqpiK]
which give
\eqn{K_{ij}=\pi_{ij}-\frac{1}{d-1}\lsp\pi\lsp\hm_{ij}\,,\qquad
\Lien\phi=\pi_\phi\,.}[eqKpi]
Then,
\eqn{\Ham_0^\hm=N\left(\pi^{ij}\pi_{ij}-\frac{1}{d-1}\lsp\pi^2 -2\lsp
\Lambda+R\right)-2\lsp N^i\nabla^j\pi_{ij}\,,\qquad
\Ham_0^\phi=\tfrac12 N(\pi_\phi^2-\partial^i\phi\lsp\partial_i\phi)
+\pi_\phi\lsp\LieN\phi\,,}[eqHamPs]
and with definitions like in \HamMom we obtain
\eqn{\begin{gathered}
  \CHam_0^\hm=\pi^{ij}\pi_{ij}-\frac{1}{d-1}\lsp\pi^2-2\lsp\Lambda
  +R\,,\qquad\CHam_0^\phi=\tfrac12(\pi_\phi^2-\partial^i\phi\lsp\partial_i
  \phi)\,,\\
  \CP_i^\hm=-2\lsp\nabla^j\pi_{ij}\,,\qquad\CP_i^\phi=\pi_\phi\lsp
  \partial_i \phi\,.
\end{gathered}}[eqHamPsTwo]
The projection of these expressions onto the boundary is trivial, and
amounts essentially to $\pi_{ij},\pi_\phi\to\pi|_{ij},\pi|_\phi$ of \Hameq,
thus translating \HamMomConstraintsTwo into constraints on the form of the
reduced classical action $S_r$.

The relations in \eqKpi can also be directly obtained by the equations
\eqn{\left\{\int\dd y'\sqrt{\hm}\,\Ham_0,\lsp\hm_{ij}(y)\right\}=
\hmdot_{ij}(y)\,,\qquad
\left\{\int\dd y'\sqrt{\hm}\,\Ham_0,\lsp\phi(y)\right\}=
\phidot(y)\,,}[]
where $\Ham_0=\Ham_0^\hm+\Ham_0^\phi$ and
\eqn{\{F(q,p),\lsp G(q,p)\}=\frac{\partial F}{\partial p}\cdot
\frac{\partial G}{\partial q}-\frac{\partial F}{\partial q}\cdot
\frac{\partial G}{\partial p}\,.}[]
The advantage of this method is that it can also be used at higher order
due to the theorem of \cite{Fukuma:2001uf}.  More specifically, from
\eqn{\left\{\int\dd y'\sqrt{\hm}\,\Ham_0,\lsp K_{ij}(y)\right\}=
\Kdot_{ij}(y)\,,\qquad
\left\{\int\dd y'\sqrt{\hm}\,\Ham_0,\lsp\Sigma(y)\right\}=
\Sigmadot(y)\,,}[]
we find that in $\HDAction$ we may use
\eqna{L_{ij}&=-\frac{1}{2\lsp(d-1)^2}\big(2\lsp(d-1)\Lambda+(d-1)R
+(d-1)\pi^{kl}\pi_{kl}-3\lsp\pi^2\big)\hm_{ij}+R_{ij}
-\frac{3}{d-1}\pi\pi_{ij}+2\lsp\pi_{i}{\!}^k\pi_{kj}\\
&\hspace{9cm}+\frac{1}{4\lsp(d-1)}\big(\partial^k\phi\lsp\partial_k\phi
-\pi_\phi^2\big)\hm_{ij}-\tfrac12\lsp\partial_i\phi\lsp\partial_j\phi\,,}[]
and
\eqn{\Lien\Lien\phi-\Liea\phi=\frac{1}{d-1}\lsp\pi\lsp\pi_\phi
-\nabla^2\phi\,.}[LienLienphi]
We notice that \LienLienphi is the ADM decomposition of the bulk equation
$\nabla^2\phi=0$. As a result, the coefficient $g$ of \BulkTerms will not
contribute to the Hamiltonian.  With our results it is straightforward to
compute
\eqna{\CHam_1^\hm&=\alpha_1\lsp\pi^i{\!}_j\pi^j{\!}_k\pi^k{\!}_l\pi^l{\!}_i
+\alpha_2\lsp\pi\pi^i{\!}_j\pi^j{}_k\pi^k{\!}_i
+\alpha_3\lsp(\pi^{ij}\pi_{ij})^2
+\alpha_4\lsp\pi^2\pi^{ij}\pi_{ij}
+\alpha_5\lsp\pi^4\\
&\quad+\beta_1\lsp\Lambda\lsp\pi^{ij}\pi_{ij}
+\beta_2\lsp\Lambda\lsp\pi^2
+\beta_3\lsp R\lsp\pi^{ij}\pi_{ij}
+\beta_4\lsp R\lsp \pi^2
+\beta_5\lsp R^{ij}\lsp\pi_i{\!}^k\pi_{kj}
+\beta_6\lsp R^{ij}\lsp\pi\pi_{ij}
+\beta_7\lsp R^{ijkl}\lsp\pi_{ik}\pi_{jl}\\
&\quad+\gamma_1\lsp \pi^{ij}\nabla_j\nabla_k\pi^{k}{\!}_i
+\gamma_2\lsp\pi^{ij}\lsp\nabla_i\partial_j\pi
+\gamma_3\lsp\pi^{ij}\lsp\nabla^2\pi_{ij}
+\gamma_4\lsp\pi\lsp\nabla^i\nabla^j\pi_{ij}
+\gamma_5\lsp\pi\lsp\nabla^2\pi\\
&\quad+\delta_1\lsp\Lambda^2
+\delta_2\lsp\Lambda R
+\delta_3\lsp R^2
+\delta_4\lsp R^{ij}R_{ij}
+\delta_5\lsp R^{ijkl}R_{ijkl}\,,}[]
and
\eqna{\CHam_1^\phi&=\epsilon_1\lsp\pi^{ij}\pi_{ij}\lsp\pi_\phi^2
+\epsilon_2\lsp\pi^2\lsp\pi_\phi^2
+\epsilon_3\lsp\pi_\phi^4\\
&\quad+\zeta_1\lsp\Lambda\lsp\pi_\phi^2
+\zeta_2\lsp R\lsp\pi_\phi^2\\
&\quad+\eta_1\lsp\pi^{ij}\lsp\pi_\phi\lsp\nabla_i\partial_j\phi
+\eta_2\lsp\pi\lsp\pi_\phi\lsp\nabla^2\phi
+\eta_3\lsp\pi^i{\!}_k\pi^{kj}\lsp\partial_i\phi\lsp\partial_j\phi
+\eta_4\lsp\pi\pi^{ij}\lsp\partial_i\phi\lsp\partial_j\phi\\
&\quad+\eta_5\lsp\pi^{ij}\pi_{ij}\lsp\partial^k\phi\lsp\partial_k\phi
+\eta_6\lsp\pi^2\lsp\partial^i\phi\lsp\partial_i\phi
+\eta_7\lsp\pi_\phi^2\lsp\partial^i\phi\lsp\partial_i\phi\\
&\quad+\theta_1\lsp\partial^i\pi_\phi\lsp\partial_i\pi_\phi
+\theta_2\lsp\pi^{ij}\lsp\partial_i\pi_\phi\lsp\partial_j\phi
+\theta_3\lsp\pi\lsp\partial^i\pi_\phi\lsp\partial_i\phi\\
&\quad+\kappa_1\lsp\Lambda\lsp\partial^i\phi\lsp\partial_i\phi
+\kappa_2\lsp R\lsp\partial^i\phi\lsp\partial_i\phi
+\kappa_3\lsp R^{ij}\lsp\partial_i\phi\lsp\partial_j\phi
+\kappa_4\lsp\nabla^2\phi\lsp\nabla^2\phi
+\kappa_5\lsp\nabla^i\partial^j\phi\lsp\nabla_i\partial_j\phi\\
&\quad+\lambda\lsp(\partial^i\phi\lsp\partial_i\phi)^2\,,
}[]
with
\begin{align}
\alpha_1&=2\lsp c\,,\qquad
\alpha_2=\frac{2}{d-1}\lsp x_5\,,\displaybreak[0]\nonumber\nonumber\\
\alpha_3&=\frac{1}{4\lsp(d-1)^2}\lsp\big(4\lsp a+(\dtwo-3\lsp d+4)\lsp b
+4\lsp (d-2)(2\lsp d-3)\lsp c-2\lsp(d-1)(d\lsp x_4+3\lsp x_5)\big)\,,\displaybreak[0]\nonumber\\
\alpha_4&=-\frac{1}{2\lsp(d-1)^3}\lsp\big(4\lsp a
+(\dtwo-3\lsp d+4)\lsp b+4\lsp(2\lsp \dtwo-5\lsp d+4)\lsp c\nonumber\\
&\hspace{2.6cm}+3\lsp d\lsp x_3-(2\lsp \dtwo-7\lsp d+2)\lsp x_4
+3\lsp(2\lsp d-1)\lsp x_5\big)\,,\displaybreak[0]\nonumber\\
\alpha_5&=\frac{1}{4\lsp(d-1)^4}\lsp\big(4\lsp a+(\dtwo-3\lsp d+4)\lsp b
+4\lsp (2 \lsp \dtwo-5\lsp d+4)\lsp c\nonumber\\
&\hspace{2.3cm}+2\lsp(3\lsp d-4)\lsp x_3-2\lsp(\dtwo-6\lsp d+6)\lsp x_4
+2\lsp(5\lsp d-6)\lsp x_5\big)\,,\displaybreak[0]\nonumber\\
\beta_1&=\frac{1}{(d-1)^2}\lsp\big(4\lsp d\lsp a - d\lsp(d-3)\lsp b
- 4\lsp(d-2)\lsp c -(d-1) (d\lsp x_4+3\lsp x_5)\big)\,,\displaybreak[0]\nonumber\\
\beta_2&=-\frac{1}{(d-1)^3}\lsp\big(4\lsp d\lsp a - d\lsp (d-3) \lsp b
- 4\lsp(d-2) \lsp c + 3 \lsp d\lsp x_3-(\dtwo-2\lsp d-2)\lsp x_4
- 3\lsp(d-2)\lsp x_5\big)\,,\displaybreak[0]\nonumber\\
\beta_3&=\frac{1}{2\lsp(d-1)^2}\lsp\big(4\lsp a + (\dtwo-3\lsp d+4)\lsp b
-4\lsp (3\lsp d-4)\lsp c -(d-1)(d\lsp x_1 + x_2 - (d-2)\lsp x_4 + 3\lsp
x_5)\big)\,,\displaybreak[0]\nonumber\\
\beta_4&=-\frac{1}{2\lsp(d-1)^3}\lsp\big(4\lsp a + (\dtwo-3\lsp d+4)\lsp b
-4\lsp(d-2)\lsp c\nonumber\\
&\hspace{2cm}-(d-1)\lsp(d-4)\lsp x_1 + 3\lsp(d-1)\lsp x_2
-3\lsp (d-2)\lsp x_3 + (\dtwo- 8\lsp d + 10)\lsp x_4
-3\lsp (3\lsp d - 4)\lsp x_5\big)\,,\displaybreak[0]\nonumber\\
\beta_5&=16\lsp c+3\lsp x_5\,,\qquad
\beta_6=\frac{2}{d-1}\lsp(x_1+2\lsp x_2-x_4-3\lsp x_5)\,,\qquad
\beta_7=-2\lsp(6\lsp c+\lsp x_2)\,,\displaybreak[0]\nonumber\\
\gamma_1&=-2\lsp(b-4\lsp c-x_2)\,,\qquad
\gamma_2=-\frac{1}{d-1}\lsp(2\lsp b+8\lsp c+2\lsp x_1+x_2)\,,\qquad
\gamma_3=-8\lsp c-x_2\,,\displaybreak[0]\nonumber\\
\gamma_4&=\frac{2}{d-1}\lsp(b-4\lsp c-x_2)\,,\qquad
\gamma_5=\frac{1}{d-1}\lsp(8\lsp c+x_2)\,,\displaybreak[0]\nonumber\\
\delta_1&=\frac{d}{(d-1)^2}\lsp\big(4\lsp d\lsp a+(d+1)\lsp b
+ 4\lsp c\big)\,,\displaybreak[0]\nonumber\\
\delta_2&=\frac{1}{(d-1)^2}\lsp\big(4\lsp d\lsp a -d\lsp(d-3)\lsp b
-4\lsp(d-2)\lsp c-(d-1)\lsp(d\lsp x_1 + x_2)\big)\,,\displaybreak[0]\nonumber\\
\delta_3&=\frac{1}{4\lsp(d-1)^2}\lsp\big(4\lsp a +(\dtwo-3\lsp d+4)\lsp b
- 4\lsp(3\lsp d-4)\lsp c + 2\lsp (d-1)\lsp((d-2)\lsp x_1 - x_2)\big)\,,
\displaybreak[0]\nonumber\\
\delta_4&=4\lsp c+x_2\,,\qquad \delta_5=c\,,
\end{align}
and
\begin{align}
\epsilon_1&=\frac{1}{4\lsp(d-1)^2}\lsp\big(4\lsp d\lsp a
-d\lsp (d-3) \lsp b - 4\lsp (d - 2)\lsp c
-(d-1)\lsp(4\lsp e - 2\lsp(d-2) \lsp f+4\lsp (d-1)\lsp h)\nonumber\\
&\hspace{2.3cm}-(d-1)(d\lsp x_4+ 3\lsp x_5)\big)\,,\displaybreak[0]\nonumber\\
\epsilon_2&=-\frac{1}{4\lsp(d-1)^3}\lsp\big(4\lsp d\lsp a
-d\lsp(d-3)\lsp b - 4\lsp(d-2)\lsp c - 2\lsp(d-1)\lsp
(2\lsp e -(d-2)\lsp f +2\lsp(d-3)\lsp h)\displaybreak[0]\nonumber\\
&\hspace{2.6cm}+3\lsp d\lsp x_3 - (\dtwo - 2\lsp d - 2)\lsp x_4
- 3\lsp(d-2)\lsp x_5\big)\,,\displaybreak[0]\nonumber\\
\epsilon_3&=\frac{d}{16\lsp(d-1)^2}\lsp\big(4\lsp d\lsp a
+(d+1)\lsp b+4\lsp c-4\lsp(d-1)\lsp(2\lsp e+f)\big)\,,\displaybreak[0]\nonumber\\
\zeta_1&=\frac{d}{2\lsp(d-1)^2}\lsp\big(4\lsp d\lsp a + (d+1)\lsp b
+ 4\lsp c -2\lsp(d-1)\lsp(2\lsp e + f)\big)\,,\displaybreak[0]\nonumber\\
\zeta_2&=\frac{1}{4\lsp(d-1)^2}\lsp\big(4\lsp d\lsp a - d\lsp(d-3)\lsp b
- 4\lsp(d-2)\lsp c - 2\lsp(d-1)\lsp(2\lsp e-(d-2)\lsp f)
-(d-1)\lsp(d\lsp x_1 + x_2) \big)\,,\displaybreak[0]\nonumber\\
\eta_1&=2(\lsp f-h)\,,\qquad
\eta_2=\frac{1}{d-1}\lsp(4\lsp h-y_3)\,,\qquad
\eta_3=-\tfrac12\lsp(8\lsp c+4\lsp h+3\lsp x_5+4\lsp y_2)\,,
\displaybreak[0]\nonumber\\
\eta_4&=\frac{1}{d-1}\lsp(4\lsp c+4\lsp h+x_4+3\lsp x_5+2\lsp y_1
+4\lsp y_2)\,,\displaybreak[0]\nonumber\\
\eta_5&=\frac{1}{4\lsp(d-1)^2}\lsp\big(4\lsp(d-2)\lsp a
-(d-2)\lsp(d-3)\lsp b + 4 (3\lsp d-4)\lsp c - 2\lsp(d-1)\lsp(2\lsp e+f)\nonumber\\
&\hspace{2.3cm}-(d-1)\lsp((d-2)\lsp x_4-3\lsp x_5 + 2\lsp d\lsp y_1
+ 2\lsp y_2)\big)\,,\displaybreak[0]\nonumber\\
\eta_6&=-\frac{1}{4\lsp(d-1)^3}\lsp\big((d-2)\lsp(4\lsp a-(d-3)\lsp b)
+ 4\lsp(3\lsp d-4)\lsp c - 2\lsp(d-1)\lsp(2\lsp e+f-4\lsp h)
\displaybreak[0]\nonumber\\
&\hspace{2.6cm}+3\lsp(d-2)\lsp x_3 -(\dtwo-8\lsp d+10)\lsp x_4
+3\lsp(3\lsp d-4)\lsp x_5 -2\lsp(d-1)\lsp((d-4)y_1-3\lsp y_2)\big)\,,\displaybreak[0]\nonumber\\
\eta_7&=\frac{1}{8\lsp(d-1)^2}\lsp\big((d-2)\lsp(4\lsp d\lsp a+(d+1)\lsp b
+4\lsp c)-2\lsp(d-1)\lsp((d-1)\lsp(4\lsp e+f)+d\lsp y_1+y_2)\big)\,,\displaybreak[0]\nonumber\\
\theta_1&=-2\lsp h+y_3\,,\qquad\
\theta_2=2\lsp(f+2\lsp h+y_2-y_3)\,,\qquad
\theta_3=-\frac{1}{d-1}\lsp(4\lsp h+2\lsp y_1+2\lsp y_2-y_3)\,,\displaybreak[0]\nonumber\\
\kappa_1&=\frac{1}{2\lsp(d-1)^2}\lsp\big((d-2)\lsp(4\lsp d\lsp a
+(d+1)\lsp b+4\lsp c) -2\lsp(d-1)\lsp(2\lsp d\lsp e+f
+d\lsp y_1+y_2)\big)\,,\displaybreak[0]\nonumber\\
\kappa_2&=\frac{1}{4\lsp(d-1)^2}\lsp\big((d-2)\lsp(4\lsp a-(d-3)\lsp b)
+4\lsp(3\lsp d-4)\lsp c -2\lsp(d-1)\lsp(2\lsp e+f)\nonumber\\
&\hspace{2.3cm}-(d-1)\lsp((d-2)\lsp x_1-x_2-2\lsp(d-2)\lsp y_1
-2\lsp y_2)\big)\,,\displaybreak[0]\nonumber\\
\kappa_3&=-\tfrac12\lsp(8\lsp c+x_2-2\lsp y_2)\,,\qquad
\kappa_4=-h+y_3\,,\qquad\kappa_5=-h\,,\displaybreak[0]\nonumber\\
\lambda&=\frac{1}{16\lsp(d-1)^2}\lsp\big(4\lsp(d-2)^2\lsp a+(5\lsp\dtwo
-15\lsp d+12)\lsp b + 4\lsp(4\lsp \dtwo - 11\lsp d + 8)\lsp c \nonumber\\
&\hspace{2.5cm}-4\lsp(d-1)\lsp(2\lsp(d-2)\lsp e+(2\lsp d-3)\lsp f
+(d-2)\lsp y_1+(2\lsp d -3)\lsp y_2)\big)\,.
\end{align}
As expected, $g$ does not appear in any of these constants. From now on we
will only consider terms quadratic in $\phi$ so that $\lambda$ above will
not be used.

With these results we can write down the equation that gives us the reduced
classical action, namely
\eqn{\CHam_r(\hm,\phi,\pi|,\pi|_\phi)\equiv\CHam_0^\hm+\CHam_0^\phi
+\CHam_1^\hm +\CHam_1^\phi=0\,.}[]
This can be written as a flow equation
\eqn{\{S_r,S_r\}+\{S_r,S_r,S_r,S_r\}=\Lag_d\,,}[FlowEq]
where the two-bracket is given by
\eqna{\hm\lsp\{S_r,S_r\}&=
(1+\beta_1\lsp\Lambda+\beta_3\lsp R)
\lsp\hm_{ik}\hm_{jl}\frac{\delta S_r}{\delta \hm_{ij}}\frac{\delta S_r
}{\delta \hm_{kl}}
-\left(\frac{1}{d-1}-\beta_2\lsp\Lambda-\beta_4\lsp R\right)
\left(\hm_{ij}\frac{\delta S_r}{\delta\hm_{ij}}\right)^2\\
&\quad+\beta_5\lsp R_{ik}\hm_{jl}\lsp\frac{\delta S_r}{\delta \hm_{ij}}\frac{\delta S_r}{\delta \hm_{kl}}
+\beta_6\lsp R_{ij}\hm_{kl}\lsp\frac{\delta S_r}{\delta
\hm_{ij}}\frac{\delta S_r}{\delta \hm_{kl}}
+\beta_7\lsp R_{ijkl}\lsp\frac{\delta S_r}{\delta \hm_{ik}}
\frac{\delta S_r}{\delta \hm_{jl}}\\
&\quad+\gamma_1\lsp\hm_{ij}\frac{\delta S_r}{\delta
\hm_{ik}}\nabla_k\nabla_l\frac{\delta S_r}{\delta \hm_{lj}}
+\gamma_2\lsp\frac{\delta S_r}{\delta \hm_{kl}}\lsp
\nabla_k\partial_l\left(\hm_{ij}\frac{\delta S_r}
{\delta \hm_{ij}}\right)
+\gamma_3\lsp\hm_{ik}\hm_{jl}\frac{\delta S_r}{\delta
\hm_{ij}}\lsp\nabla^2\frac{\delta S_r}{\delta \hm_{kl}}\\
&\quad+\gamma_4\lsp\hm_{ij}\frac{\delta S_r}{\delta
\hm_{ij}}\lsp\nabla_k\nabla_l\frac{\delta S_r}{\delta \hm_{kl}}
+\gamma_5\lsp\hm_{ij}\hm_{kl}\frac{\delta S_r}{\delta
\hm_{ij}}\lsp\nabla^2\frac{\delta S_r}{\delta \hm_{kl}}\\
&\quad+\big(\tfrac12+\zeta_1\lsp\Lambda+\zeta_2\lsp R\big)
\left(\frac{\delta S_r}{\delta\phi}\right)^2\\
&\quad+\eta_1\lsp\frac{\delta S_r}{\delta \hm_{ij}}\lsp\frac{\delta
S_r}{\delta\phi}\lsp\nabla^i\partial^j\phi
+\eta_2\lsp\hm_{ij}\frac{\delta S_r}{\delta\hm_{ij}}
\lsp\frac{\delta S_r}{\delta\phi}\lsp\nabla^2\phi
+\eta_3\lsp\hm_{ik}\frac{\delta S_r}{\delta\hm_{ij}}
\frac{\delta S_r}{\delta\hm_{kl}}\lsp\partial_j\phi\lsp\partial_l\phi\\
&\quad+\eta_4\lsp\hm_{ij}\frac{\delta S_r}{\delta\hm_{ij}}
\frac{\delta S_r}{\delta\hm_{kl}}\lsp\partial_k\phi\lsp\partial_l\phi
+\eta_5\lsp\hm_{ik}\hm_{jl}\frac{\delta S_r}{\delta\hm_{ij}}\frac{\delta
S_r}{\delta\hm_{kl}}\lsp\partial^m\phi\lsp\partial_m\phi\\
&\quad+\eta_6\left(\hm_{ij}
\frac{\delta S_r}{\delta\hm_{ij}}\right)^2\lsp
\partial^k\phi\lsp\partial_k\phi
+\eta_7\left(\frac{\delta
S_r}{\delta\phi}\right)^2\partial^i\phi\lsp\partial_i\phi\\
&\quad+\theta_1\lsp\partial^i\frac{\delta S_r}{\delta\phi}\lsp\partial_i\frac{\delta S_r}{\delta\phi}
+\theta_2\lsp\frac{\delta S_r}{\delta\hm_{ij}}\lsp
\partial_i\frac{\delta S_r}{\delta\phi}\lsp\partial_j\phi
+\theta_3\lsp\hm_{ij}\frac{\delta S_r}{\delta\hm_{ij}}\lsp
\partial^k\frac{\delta S_r}{\delta\phi}\lsp\partial_k\phi\,,}[TwoS]
the four-bracket is given by
\eqna{\hm^2\lsp\{S_r,S_r,S_r,S_r\}&=
\alpha_1\lsp\hm_{iq}\hm_{jk}\hm_{lm}\hm_{np}
\frac{\delta S_r}{\delta\hm_{ij}}
\frac{\delta S_r}{\delta\hm_{kl}}
\frac{\delta S_r}{\delta\hm_{mn}}
\frac{\delta S_r}{\delta\hm_{pq}}\\
&\quad+\alpha_2\lsp\hm_{ij}
\frac{\delta S_r}{\delta\hm_{ij}}
\hm_{kq}\hm_{lm}\hm_{np}
\frac{\delta S_r}{\delta\hm_{kl}}
\frac{\delta S_r}{\delta\hm_{mn}}
\frac{\delta S_r}{\delta\hm_{pq}}
+\alpha_3\left(
\hm_{ik}\hm_{jl}\frac{\delta S_r}{\delta\hm_{ij}}
\frac{\delta S_r}{\delta\hm_{kl}}\right)^2\\
&\quad+\alpha_4\left(\hm_{ij}\frac{\delta
S_r}{\delta\hm_{ij}}\right)^2
\hm_{km}\hm_{ln}\frac{\delta S_r}{\delta\hm_{kl}}
\frac{\delta S_r}{\delta\hm_{mn}}
+\alpha_5\left(\hm_{ij}\frac{\delta S_r}{\delta\hm_{ij}}\right)^4\\
&\quad+\epsilon_1\lsp\lsp\hm_{ik}\hm_{jl}
\frac{\delta S_r}{\delta\hm_{ij}}\frac{\delta S_r}{\delta\hm_{kl}}
\left(\frac{\delta S_r}{\delta\phi}\right)^2
+\epsilon_2\left(\hm_{ij}\frac{\delta S_r}{\delta\hm_{ij}}\right)^2
\left(\frac{\delta S_r}{\delta\phi}\right)^2
+\epsilon_3\left(\frac{\delta S_r}{\delta\phi}\right)^4\,,}[FourS]
and
\eqna{\Lag_d&=2\lsp\Lambda-R-\delta_1\lsp\Lambda^2-\delta_2\lsp\Lambda\lsp
R -\delta_3\lsp R^3-\delta_4\lsp R^{ij}R_{ij}
-\delta_5\lsp R^{ijkl}R_{ijkl}\\
&\quad+\tfrac12\lsp\partial^i\phi\lsp\partial_i\phi
-\kappa_1\lsp\Lambda\lsp\partial^i\phi\lsp\partial_i\phi
-\kappa_2\lsp R\lsp\partial^i\phi\lsp\partial_i\phi
-\kappa_3\lsp R^{ij}\lsp\partial_i\phi\lsp\partial_j\phi
-\kappa_4\lsp\nabla^2\phi\lsp\nabla^2\phi
-\kappa_5\lsp\nabla^i\partial^j\phi\lsp\nabla_i\partial_j\phi\,.}[Lagd]

\newsec{Trace anomaly from the flow equation}[secTA]
The flow equation \FlowEq becomes useful if we make an ansatz for $S_r$
following \cite{deBoer:1999xf, deBoer:2000cz}:
\eqn{\frac{1}{2\lsp\tilde{\kappa}_{d+1}^{\lsp 2}}S_r[\hm,\phi]=
\frac{1}{2\lsp\tilde{\kappa}_{d+1}^{\lsp 2}}S_{\text{loc}}[\hm,\phi]
+\Gamma[\hm,\phi]\,,}[SrSplit]
where $2\lsp\tilde{\kappa}_{d+1}^{\lsp 2}=16\lsp\pi\lsp G_{d+1}$ with
$G_{d+1}$ the $(d+1)$-dimensional Newton constant, $\Gamma$ is the
generating functional of the boundary field theory, and $S_{\text{loc}}$
contains local counterterms. Equation \SrSplit is only valid close to the
boundary. Contributions to $S_{\text{loc}}$ are classified according to
their scaling behavior close to the boundary. This is described by an
appropriately defined weight $w$,\foot{See section \ref{secOsb} below for
more details on the weight.} giving rise to the relation
\eqn{S_{\text{loc}}=\int\dd y\sqrt{\hm}\,\Lag_{\text{loc}}\,,\qquad
\Lag_{\text{loc}}=\sum_{w=0,2,\ldots}\Lag_{\text{loc}}^{(w)}\,.}[Sbyweight]
A derivative has $w=1$, and so a curvature has $w=2$. We now use \Sbyweight
in \FlowEq and obtain independent equations for every weight. For $w=0,2$
we find
\eqna{\{\Sloc,\Sloc\}_{w=0} +\{\Sloc,\Sloc,\Sloc,\Sloc\}_{w=0}&=
(2-\delta_1\lsp\Lambda)\Lambda\,,\\
\{\Sloc,\Sloc\}_{w=2} +\{\Sloc,\Sloc,\Sloc,\Sloc\}_{w=2}&=
-(1+\delta_2\lsp\Lambda)R+(\tfrac12-\kappa_1\lsp\Lambda)
\partial^i\phi\lsp\partial_i\phi\,,}[]
respectively, which, with
\eqn{\Lambda=-\frac{d\lsp(d-1)}{2\lsp\ell^2}+\frac{d\lsp(d-3)}{2\lsp\ell^4}\big(d\lsp(d+1)\lsp a+d\lsp b+2\lsp c\big)\,,}[eqCC]
so that we have AdS space with radius $\ell$ asymptotically, allow us to
determine
\eqn{\Lagloc^{(0)}=W\,,\qquad\Lagloc^{(2)}=-\Phi\lsp R+\tfrac12\lsp M\lsp
\partial^i\phi\lsp\partial_i\phi\,,}[]
with
\eqna{W&=-\frac{2\lsp(d-1)}{\ell}-\frac{1}{\ell^3}\big(4\lsp d\lsp(d+1)\lsp
a+4\lsp d\lsp b+8\lsp c+d\lsp(\dtwo x_3+d\lsp x_4+x_5)\big)\,,\\
\Phi&=\frac{\ell}{d-2}-\frac{1}{\ell}\left(\frac{2}{(d-1)(d-2)}\big(d\lsp
(d+1)\lsp a+d\lsp b+2\lsp c\big)-d\lsp x_1-x_2
-\frac{3}{2\lsp(d-1)}(\dtwo x_3+d\lsp x_4+x_5)\right),\\
M&=\frac{\ell}{d-2}+\frac{1}{\ell}\left(\frac{2}{d-1}\big(d\lsp(d+1)\lsp a
+d\lsp b+2\lsp c\big)-\frac{2}{d-2}(d\lsp(d+1)\lsp e+d\lsp f-2\lsp
h)\right.\\
&\hspace{7.2cm}\left.+\frac{3}{2\lsp(d-1)}(\dtwo x_3+d\lsp x_4+x_5) -2\lsp
(d\lsp y_1+\lsp y_2)\right)\,,}[WPhiM]
where we choose the negative sign for the $1/\ell$ term of $W$ so that the
$\ell$ term of $M$ is positive. Note that we are only able to determine
$W,\Phi$ and $M$ up to specific powers of $\ell$ as seen in \WPhiM,
consistently with the terms included in \BulkTerms and \BoundaryTerms.
Further terms in the $\ell$-expansion of $W,\Phi$ and $M$ depend generally
also on even higher-derivative terms than the ones considered in \BulkTerms
and \BoundaryTerms.

\subsec{Four-dimensional trace anomaly}
At weight four the four-dimensional trace anomaly can be evaluated using
the definition
\eqn{\langle T^i{\!}_i\rangle=-\frac{2}{\sqrt{\hm}}\hm_{ij}
\frac{\delta\Gamma}{\delta \hm_{ij}}\,.}[defTraceT]
We assign weight four to $\delta\Gamma/\delta\hm_{ij}$ and
$\delta\Gamma/\delta\phi$, and then \FlowEq gives\foot{In
\eqref{weightfour} by $2\lsp\{\Sloc,\Gamma\}$ we mean
$\{\Sloc,\Gamma\}+\{\Gamma,\Sloc\}$ and similarly for
$4\lsp\{\Sloc,\Sloc,\Sloc,\Gamma\}$.}
\eqna{2\lsp\{\Sloc,\Gamma\}_{w=4}+4\{\Sloc,\Sloc,\Sloc,\Gamma\}_{w=4}=
-\smash{\frac{1}{2\lsp \kappa_5^2}}&\big(\{\Sloc,\Sloc\}_{w=4}
+\{\Sloc,\Sloc,\Sloc,\Sloc\}_{w=4}\\
&\hspace{-3.5cm}+\delta_3\lsp R^2+\delta_4\lsp R^{ij}R_{ij}+\delta_5\lsp R^{ijkl}R_{ijkl}\\
&\hspace{-3.5cm}+\kappa_2\lsp R\lsp\partial^i\phi\lsp\partial_i\phi
+\kappa_3\lsp R^{ij}\lsp\partial_i\phi\lsp\partial_j\phi
+\kappa_4\lsp\nabla^2\phi\lsp\nabla^2\phi
+\kappa_5\lsp\nabla^i\partial^j\phi\lsp\nabla_i\partial_j\phi\big)\,.
}[weightfour]
It turns out that with $W$ as in \WPhiM the left-hand side of \weightfour
does not contain a $1/\ell^3$ contribution for any $d$, and so it is
simply equal to $(1/\ell)\lsp\langle T^i{\!}_i\rangle$ at the order we're
working in. It is also straightforward to work out the right-hand side of
\weightfour, and using \WPhiM we finally find
\eqna{\langle T^i{\!}_i\rangle=\smash{\frac{\ell}{2\lsp
\tilde{\kappa}_5^2}}&\smash{\Big(}-\tfrac12\lsp\big(\tfrac14\lsp\ell^2
-10\lsp a-2\lsp b-c\big)\lsp E_4
+\tfrac12\lsp\big(\tfrac14\lsp \ell^2-10\lsp a-2\lsp b+c\big)\lsp
W^{ijkl}W_{ijkl}\\
&\;\,\,+\tfrac12\big(\tfrac14\lsp\ell^2-10\lsp e-2\lsp f+2\lsp h\big)
\lsp\big(\nabla^2\phi\lsp\nabla^2\phi -2\lsp R^{ij}\partial_i\phi\lsp
\partial_j\phi+\tfrac23\lsp R\lsp\partial^i\phi\lsp\partial_i\phi\big)
\Big)\,,}[FourdTA]
where the Euler term is
\eqn{E_4=R^{ijkl}R_{ijkl}-4\lsp R^{ij}R_{ij}+R^2\,,}[Efour]
and the Weyl tensor is here the $d=4$ version of
\eqn{W_{ijkl}=R_{ijkl}+
\tfrac{2}{d-2}(\hm_{i[l}R_{k]j}+\hm_{j[k}R_{l]i})
+\tfrac{2}{(d-1)(d-2)}\hm_{i[k}\hm_{l]j}R\,.}[Weyld]
In \FourdTA the part of the anomaly quadratic in $\phi$ is in accord with
the $d=4$ version of the Paneitz operator~\cite{Paneitz} (see also
\cite{Fradkin:1981jc, Fradkin:1982xc, Fradkin:1983tg, Riegert:1984kt}).
Note that although present throughout the calculation the constants
$x_1,\ldots,x_5,y_1,y_2,y_3$ do not contribute to the final result
\FourdTA.

\subsec{Six-dimensional trace anomaly}
To compute the $d=6$ anomaly we have to consider the weight-six part of the
flow equation \FlowEq. At weight four we set
\eqn{\Lagloc^{(4)}=XR^2+YR^{ij}R_{ij}+ Z\lsp R^{ijkl}R_{ijkl}
+V R\lsp\lsp \partial^i\phi\lsp\partial_i\phi+U\lsp
R^{ij}\lsp\partial_i\phi\lsp\partial_j\phi
+T\lsp\lsp\nabla^2\phi\lsp\nabla^2\phi\,,}[]
and the weight-four part of \FlowEq allows us to determine
\begin{align}
X&=\frac{d\lsp\ell^3}{4\lsp(d-1)(d-2)^2(d-4)}
-\frac{\ell}{2\lsp(d-1)^2(d-2)^2(d-4)}\big(3\lsp \dtwo(d+1)\lsp a
-3\lsp\dtwo b+2\lsp(2\lsp\dtwo-3\lsp d+4)\lsp c\big)\nonumber\\
&\quad-\frac{\ell}{2\lsp(d-1)}\left(x_1-\frac{1}{d-2}\lsp x_2
+\frac{1}{4\lsp(d-1)(d-2)^2}\left(3\lsp d\lsp(\dtwo-8\lsp d+8)\lsp
x_3\right.\right.\nonumber\\
&\hspace{9cm}\left.\vphantom{\frac{d}{d-1}}\left.-(5\lsp\dtwo+8\lsp
d-16)\lsp x_4 -3\lsp(7\lsp d-8)\lsp x_5\right)\right),\displaybreak[0]\nonumber\\
Y&=-\frac{\ell^3}{(d-2)^2(d-4)}+\frac{2\lsp\ell}{(d-1)(d-2)^2(d-4)}
\big(3\lsp d\lsp(d+1)\lsp a+3\lsp d\lsp b+2\lsp(\dtwo-3\lsp d+5)\lsp
c\big)\nonumber\\
&\quad-\frac{\ell}{d-2}\left(x_2+\frac{1}{2\lsp(d-1)(d-2)}\left(3\lsp\dtwo
x_3+d\lsp(2\lsp d+1)\lsp x_4+3\lsp(2\lsp d+1)\lsp x_5\right)\right),
\displaybreak[0]\nonumber\\
Z&=-\frac{\ell}{d-4}\lsp c\,,\displaybreak[0]\nonumber\\
V&=-\frac{d\lsp\ell^3}{4\lsp(d-1)(d-2)^2(d-4)}-\frac{d\lsp\ell}
{2\lsp(d-1)^2(d-2)^2}\big(d\lsp(d+1)\lsp a +d\lsp b+2\lsp c\big)\nonumber\\
&\quad+\frac{\ell}{2\lsp(d-1)(d-2)^2(d-4)}\big(\dtwo(d+1)\lsp e+\dtwo
f-2\lsp(3\lsp d-4)\lsp h \big)\nonumber\\
&\quad+\frac{\ell}{4\lsp(d-1)}\left(x_1-\frac{1}{d-2}\lsp x_2
+\frac{1}{2\lsp(d-1)(d-2)^2}\left(3\lsp d\lsp(\dtwo-8\lsp d+8)\lsp
x_3\right.\right.\nonumber\\
&\hspace{9cm}\left.\vphantom{\frac{d}{d-1}}\left.-(5\lsp\dtwo+8\lsp
d-16)\lsp x_4 -3\lsp(7\lsp d-8)\lsp x_5\right)\right)\nonumber\\
&\quad-\frac{\ell}{2\lsp(d-1)}\left(y_1-\frac{1}{d-2}\lsp y_2\right),
\displaybreak[0]\nonumber\\
U&=\frac{\ell^3}{(d-2)^2(d-4)}+\frac{2\lsp\ell}{(d-1)(d-2)^2}
\big(d\lsp(d+1)\lsp a+d\lsp b+2\lsp c\big)\nonumber\\
&\quad-\frac{\ell}{(d-2)^2(d-4)}\big(2\lsp d\lsp(d+1)\lsp e+2\lsp d\lsp f
+(\dtwo-8\lsp d+8)\lsp h\big)\nonumber\\
&\quad+\frac{\ell}{2\lsp(d-2)}\left(x_2+
\frac{1}{(d-1)(d-2)}\left(3\lsp\dtwo x_3+d\lsp(2\lsp d+1)\lsp x_4
+3\lsp(2\lsp d-1)\lsp x_5\right)\right)-\frac{\ell}{d-2}\lsp y_2\,,
\displaybreak[0]\nonumber\\
T&=-\frac{\ell^3}{2\lsp(d-2)^2(d-4)}-\frac{\ell}{(d-1)(d-2)^2}\big(d\lsp
(d+1)\lsp a +d\lsp b+2\lsp c\big)\nonumber\\
&\quad+\frac{\ell}{(d-2)^2(d-4)}\big(d\lsp(d+1)\lsp e+d\lsp f+\dtwo
g+(\dtwo-9\lsp d+16)\lsp h\big)\nonumber\\
&\quad-\frac{3\lsp\ell}{4\lsp(d-1)(d-2)^2}\big(\dtwo x_3+d\lsp x_4+x_5
\big)-\frac{\ell}{d-2}\lsp y_3\,.
\end{align}
With these results as well as \WPhiM and the weight-six part of \FlowEq we
can finally get the holographic trace anomaly in $d=6$:
\eqna{\langle T^i{\!}_i\rangle&=\smash{\frac{\ell^3}{2\lsp\tilde{\kappa}_7^2}
\Big(}\tfrac{1}{48}\big(\tfrac14\ell^2-21\lsp a-3\lsp b-c\big)E_6
-\tfrac{1}{4}\big(\tfrac14\ell^2-21\lsp a-3\lsp b+\tfrac13\lsp c\big)I_1\\
&\hspace{1.45cm}-\tfrac{1}{16}\big(\tfrac14\ell^2-21\lsp a-3\lsp b
-\tfrac73\lsp c\big)I_2
+\tfrac{1}{48}\big(\tfrac14\ell^2-21\lsp a-3\lsp b+3\lsp c\big)I_3\\
&\hspace{1.45cm}+\tfrac{1}{48}\big(\tfrac14\ell^2-21\lsp a-3\lsp b+3\lsp c\big)J_1
+\tfrac{1}{6}\big(\tfrac14\ell^2-21\lsp a-3\lsp b+3\lsp c\big)J_2\\
&\hspace{1.45cm}-\tfrac{1}{8}\big(\tfrac14\ell^2-21\lsp a-3\lsp b+3\lsp
c\big)J_3
+\tfrac{1}{48}\big(\tfrac14\ell^2-21\lsp a-3\lsp b+3\lsp c\big)J_4\\
&\hspace{1.45cm}-\tfrac{1}{32}\big(\tfrac14\ell^2-21\lsp e-3\lsp f
+3\lsp h\big)\big(\partial^i\nabla^2\phi\lsp\partial_i\nabla^2\phi
-4\lsp R^{ij}\lsp\nabla_i\partial_j\phi\lsp\nabla^2\phi
+R\lsp\lsp\nabla^2\phi\lsp\nabla^2\phi\\
&\hspace{6.2cm}+2\lsp(2\lsp R^{ikjl}R_{kl}+R^{ik}R^{j}{\!}_k -R\lsp R^{ij}
+\nabla^2 R^{ij})\partial_i\phi\lsp\partial_j\phi\\
&\hspace{6.2cm}-(R^{ij}R_{ij}-\tfrac{9}{25}\lsp R^2+\tfrac35\lsp \nabla^2
R)\lsp\partial^k\phi\lsp\partial_k\phi\big)\\
&\hspace{1.45cm}-\tfrac14\lsp c\lsp\lsp W^{iklm}W^j{\!}_{klm}\lsp
\partial_i\phi\lsp\partial_j\phi
+\tfrac{1}{20}\lsp c\lsp\lsp W^{ijkl}W_{ijkl}\lsp
\partial^m\phi\lsp\partial_m\phi\Big)\,.}[SixdTA]
Here $E_6$ is the Euler term in six dimensions and $I_{1,2,3}$ the three
terms with Weyl-invariant densities. Explicit expressions for these, as
well as for the trivial anomalies $J_{1,\ldots,4}$, are given in Appendix
\ref{appSixdCurv}. The first three lines in the part quadratic in $\phi$
are in accord with the Branson operator \cite{Branson}, while the next two
terms involving the Weyl tensor were shown to appear generally in CFTs in
six dimensions in~\cite{Osborn:2015rna}. Note that total derivatives have
been dropped in \SixdTA. Just like in \FourdTA the constants
$x_1,\ldots,x_5,y_1,y_2,y_3$ do not contribute to the final result \SixdTA.

We note here that the results \FourdTA and \SixdTA have an obvious
generalization to the case where $\phi\to\phi^a$, where $a$ is a flavor
index.

\subsec{Bounds}[bounds]
From the field theory point of view there are bounds derived by requiring
positivity of the energy flux in lightlike directions \cite{Hofman:2008ar}.
These are bounds on the three-point function of the stress-energy tensor,
which in $d=6$ take the form \cite{deBoer:2009pn}
\eqna{C_1&\equiv1-\tfrac15 t_2-\tfrac{2}{35}t_4\ge0\,,\qquad
C_2\equiv1-\tfrac15 t_2-\tfrac{2}{35}t_4+\tfrac12 t_2\ge0\,,\\
C_3&\equiv1-\tfrac15 t_2-\tfrac{2}{35}t_4+\tfrac45(t_2+t_4)\,,}[Cineqs]
where $t_2$ and $t_4$ correspond to the angular dependencies of the energy
flux at null infinity. They are related to the coefficients $c_1$, $c_2$
and $c_3$ of the two- and three-point function of the stress-energy tensor
by
\eqn{t_2=\frac{15\lsp(23\lsp c_1-44\lsp c_2+144\lsp c_3)}{16\lsp c_3}\,,
\qquad
t_4=-\frac{105\lsp(c_1-2\lsp c_2+6\lsp c_3)}{2\lsp c_3}\,,}[tres]
where we use results for free fields first obtained in
\cite{Bastianelli:2000hi} (see also \cite{deBoer:2009pn, Osborn:2015rna}).
The coefficient $c_3$ appears in the two-point function of the
stress-energy tensor and thus $c_3>0$.

Our computation \SixdTA allows us to determine
\eqna{c_1&=-\frac{1}{4800\sqrt{10}}\frac{L^5}{2\tilde{\kappa}_7^2}
\big(5+\sqrt{25-60\lsp z}\big)^{3/2}\big(3\lsp(5-20\lsp z
+\sqrt{25-60\lsp z})+160\lsp \tilde{c}\big)\,,\\
c_2&=-\frac{1}{19\lsp200\sqrt{10}}\frac{L^5}{2\tilde{\kappa}_7^2}
\big(5+\sqrt{25-60\lsp z}\big)^{3/2}\big(3\lsp(5-20\lsp z
+\sqrt{25-60\lsp z})-160\lsp \tilde{c}\big)\,,\\
c_3&=\frac{1}{19\lsp200\sqrt{10}}\frac{L^5}{2\tilde{\kappa}_7^2}
\big(5+\sqrt{25-60\lsp z}\big)^{3/2}\big(5-20\lsp z
+\sqrt{25-60\lsp z}+160\lsp \tilde{c}\big)\,,}[holcs]
where
\eqn{L^2=-15/\Lambda\,,\qquad
z=42\lsp\tilde{a}+6\lsp \tilde{b}+2\lsp \tilde{c}\,,\qquad
\tilde{a}=a/L^2\,,\quad
\tilde{b}=b/L^2\,,\quad\tilde{c}=c/L^2\,.}[zabctilde]
We note that \holcs are invariant under field redefinitions.\foot{With
Einstein or Lovelock gravity in the bulk $c_1$, $c_2$ and $c_3$ in \holcs
are such that $t_4=0$. This implies that the corresponding boundary theory
is superconformal~\cite{deBoer:2009pn}.} With these results we can now use
\Cineqs with \tres to obtain
\eqn{-\frac{1}{880}\big(5-20\lsp z+\sqrt{25-60\lsp z}\big)\le\tilde{c}
\le\frac{1}{80}\big(5-20\lsp z+\sqrt{25-60\lsp z}\big)\,,\qquad
z\le\frac{7}{20}\,.}[czbound]
As we see $\tilde{c}$ can take both negative and positive values. For
Gauss--Bonnet gravity in the bulk we reproduce the result of
\cite{deBoer:2009pn, Camanho:2009vw, Buchel:2009sk}
($\tilde{c}\to\lambda/12$ in their notation),
\eqn{-\frac{5}{192}\le\tilde{c}\le\frac{1}{64}\,.}[]

\newsec{Trace anomaly away from fixed points}[secOsb]
In this section we establish a connection with Osborn's local RG. We will
work in Einstein gravity with scalar fields $\phi^a$. Since we are now
interested in the flow of the boundary theory we will keep the
$\phi$-dependence of the various quantities that enter our expressions. Our
flow equation is now
\eqn{\{S_r,S_r\}=\Lag_d\,,}[FlowEqII]
with
\eqna{\hm\lsp\{S_r,S_r\}&=\hm_{ik}\hm_{jl}\frac{\delta S_r}
{\delta\hm_{ij}}\frac{\delta S_r}{\delta\hm_{kl}}
-\frac{1}{d-1}\left(\hm_{ij}\frac{\delta S_r}{\delta\hm_{ij}}\right)^2
+\tfrac12\lsp H^{ab}(\phi)\lsp\frac{\delta S_r}{\delta\phi^a}
\frac{\delta S_r}{\delta\phi^b}\,,\\
\Lag_d&=V(\phi)-R+\tfrac12\lsp H_{ab}(\phi)\lsp\partial^i\phi^a\lsp
\partial_i\phi^b\,.}[bracketEin]

Following the prescription of \cite{deBoer:1999xf}, we use the splitting
\SrSplit and write down the local terms up to second order in derivatives:
\eqn{\Lagloc=W(\phi)-\Phi(\phi)\lsp R
+\tfrac12\lsp M_{ab}(\phi)\lsp
\partial^i\phi^a\lsp\partial_i\phi^b\,.}[SUGRAcounter]
Using \SUGRAcounter we collect terms of the same functional form in
\FlowEqII and find
\fourseqn{V&=\tfrac12\lsp H^{ab}\lsp\partial_a W\lsp\partial_b W
-\frac{d}{4\lsp(d-1)}W^2\,,}[wZeroTwoEqI]
{-1&=\frac{d-2}{2\lsp(d-1)}W\Phi
-H^{ab}\lsp\partial_a W\lsp\partial_b\Phi\,,}[wZeroTwoEqII]
{\tfrac12\lsp H_{ab}&=-\frac{d-2}{4\lsp(d-1)}WM_{ab}
-W\partial_a\partial_b\Phi
-\Gamma^M_{abc}\lsp\partial_d W\lsp H^{cd}\,,}[wZeroTwoEqIII]
{0&=W\lsp\partial_a\Phi+M_{ab}\lsp\partial_c W\lsp H^{bc}\,,
}[wZeroTwoEqIV][wZeroTwoEqs]
where $\Gamma^M_{abc}=\tfrac12(\partial_aM_{bc} +\partial_b
M_{ac}-\partial_c M_{ab})$ and $\partial_a=\partial/\partial\phi^a$.  The
holographic beta function is given by
\eqn{\beta^a=-2\lsp(d-1)\lsp H^{ab}\lsp\partial_b \log W\,,}[holbeta]
where, although $W$ is negative\foot{We choose $W<0$ so that $H_{ab}$ and
$M_{ab}$ can be taken positive-definite consistently with \wZeroTwoEqIII in
the limit where the $\phi$ dependence is neglected. In that case $\Phi$ is
positive as can be seen from \wZeroTwoEqII.} and dimensionful, we use
$\partial_a\log W$ for $\partial_a W/W$.

In order to establish a connection with Osborn's local RG, we have to study
the scaling behavior of the scalar fields $\phi$, which are viewed as
sources in the dual field theory. This is done by introducing a mass term
in the bulk potential. For convenience we use $\phi$ for the massless bulk
scalar fields, and $\chi$ for the massive ones with mass $m_\chi$. By
solving \wZeroTwoEqI perturbatively to second order in $\chi$, one can
obtain $\Delta_{\chi}=\frac{1}{2}d+\sqrt{\frac{1}{4}\dtwo+
m_{\chi}^2\ell^2}$~\cite{deBoer:1999xf}.  This reproduces the standard
relation between the mass $m_{\chi}$ of the scalar field $\chi$ and the
scaling dimension $\Delta_{\chi}$ of the dual operator. To obtain the trace
anomaly from the flow equation \FlowEqII, we have to assign ``weight'' zero
to the scalar fields $\phi$, and weight $d-\Delta_{\chi}$ to $\chi$. This
can be shown to be equivalent to the prescription of \cite{deBoer:2000cz},
which uses a scaling argument.

\subsec{Marginal operators}
Here we only include massless scalar fieds $\phi$, which correspond to
marginal operators in the dual field theory. In this case, $\Sloc$ nicely
breaks down to separate contributions of weight zero and two, with
\eqn{\Lagloc^{(0)}=W(\phi)\,,\qquad\Lagloc^{(2)}=-\Phi(\phi)\lsp R
+\tfrac12\lsp M_{ab}(\phi)\lsp
\partial^i\phi^a\lsp\partial_i\phi^b\,.}[]
The weight-$d$ part of the flow equation \FlowEqII can be brought to the
form
\eqn{-\frac{2}{\sqrt{\hm}}\lsp\hm_{ij}\frac{\delta\Gamma}{\delta\hm_{ij}}
-\beta^a\frac{1}{\sqrt{\hm}}\frac{\delta\Gamma}{\delta\phi^a}=
-\frac{1}{2\lsp\tilde{\kappa}_{d+1}^2}\frac{2\lsp(d-1)}{W}\lsp
\{\Sloc,\Sloc\}_{w=d}\,.}[wdEq]
The left-hand side of \wdEq is clearly $\langle T^i{\!}_i\rangle
-\beta^a\langle\CO_a\rangle$, where $\CO_a$ are marginal operators. The
right-hand side of \wdEq can be easily computed using \bracketEin. It is
clear that \wdEq is the holographic counterpart of the local RG equation of
Osborn \cite{Osborn:1991gm}. We should note here that \wdEq involves bare
quantities, but as was shown already in~\cite{deBoer:1999xf} we can
essentially write down the same equation with the renormalized quantities.

Osborn's expression is the starting point for the derivation of Weyl
consistency conditions that include an equation that resembles an
$a$-theorem. Here we will compute the various quantities that enter
Osborn's expression holographically, focusing in the four-dimensional case.
Using \bracketEin we compute
\eqna{\{S_\text{loc},S_\text{loc}\}_{w=4}&=-\tfrac12\lsp\Phi^2\lsp E_4
+\tfrac12\lsp\Phi^2\lsp W^{ijkl}W_{ijkl}
+\tfrac12\lsp H^{ab}\lsp\partial_a\Phi\lsp\partial_b\Phi\lsp R^2\\
&\quad-2\lsp\Phi\lsp\partial_a\Phi\lsp
G^{ij}\lsp\nabla_i\partial_j\phi^a
+M_{ab}\lsp\partial_c\Phi\lsp H^{bc}\lsp R\lsp\nabla^2\phi^a\\
&\quad-\Phi\lsp(M_{ab}+2\lsp\partial_a\partial_b\Phi)\lsp
G^{ij}\lsp\partial_i\phi^a\lsp\partial_j\phi^b
-\tfrac16\lsp(\Phi M_{ab}-6\lsp\Gamma^M_{abc}\lsp\partial_d \Phi\lsp
H^{cd})\lsp R\lsp\partial^i\phi^a\lsp\partial_i\phi^b\\
&\quad+\tfrac12\lsp(M_{ac}M_{bd}H^{cd}
-2\lsp\partial_a\Phi\lsp\partial_b\Phi)\lsp
\nabla^2\phi^a\lsp\nabla^2\phi^b
+\partial_a\Phi\lsp\partial_b\Phi\lsp\nabla^i\partial^j
\phi^a\lsp\nabla_i\partial_j\phi^b\\
&\quad-\tfrac12\lsp(M_{ab}\lsp\partial_c\Phi
+4\lsp\partial_a\partial_b\Phi\lsp\partial_c\Phi
-2\lsp\Gamma^M_{abd}M_{ce}H^{de})\lsp
\partial^i\phi^a\lsp\partial_i\phi^b\lsp\nabla^2\phi^c\\
&\quad+\lsp(M_{ab}\lsp\partial_c\Phi
+2\lsp\partial_a\partial_b\Phi\lsp\partial_c\Phi)
\partial^i\phi^a\lsp\partial^j\phi^b\lsp\nabla_i\partial_j\phi^c\\
&\quad-(\tfrac{1}{12}\lsp M_{ab}M_{cd}-\tfrac14\lsp M_{ac}M_{bd}
+\tfrac12\lsp\Gamma^M_{abe}\Gamma^M_{cdf}H^{ef}\\
&\qquad\,\,\,+\tfrac12\lsp M_{ab}\lsp\partial_c\partial_d\Phi
-M_{ac}\lsp\partial_b\partial_d\Phi
+\partial_a\partial_b\Phi\lsp \partial_c\partial_d\Phi
-\partial_a\partial_c\Phi\lsp\partial_b\partial_d\Phi)\lsp
\partial^i\phi^a\lsp\partial_i\phi^b\lsp\partial^j\phi^c\lsp
\partial_j\phi^d\,,}[]
where $G_{ij}$ is the Einstein tensor.

To match with Osborn's expression we write
\eqn{(\DeltaW_\sigma-\Delta_\sigma^\beta)\lsp\Gamma=-
\int\dfour y\sqrt{\hm}\,\sigma\lsp
\frac{1}{2\lsp\tilde{\kappa}_{5}^2}\frac{6}{W}\lsp
\{S_\text{loc},S_\text{loc}\}_{w=4}\,,}[HolOsbEq]
where, with the definitions
\eqn{\DeltaW_\sigma=-2\int\dfour
y\,\sigma\lsp\hm_{ij}\frac{\delta}{\delta\hm_{ij}}\,,\qquad
\Delta_\sigma^\beta=\int\dfour
y\,\sigma\lsp\beta^a\frac{\delta}{\delta\phi^a}\,,}[]
we have
\eqn{\DeltaW_\sigma\Gamma=\int\dfour y\sqrt{\hm}\,\sigma\lsp\langle
T^i{\!}_i\rangle\,,\qquad \Delta_\sigma^\beta\Gamma=\int\dfour y\sqrt{\hm}
\,\sigma\lsp\beta^a\langle\CO_a\rangle\,.}[]
The general form of the anomaly is
\eqna{(\DeltaW_\sigma -\Delta_\sigma^\beta)\lsp\Gamma&=-\smash{\int}\dfour
y\sqrt{\hm}\,\sigma\lsp\big(A\lsp E_4+B\lsp R^2-C\lsp W^{ijkl}W_{ijkl}\\
&\hspace{1cm}+\tfrac13\lsp E^{\smash{\phi}}_a\lsp\partial^i\phi^a
\lsp\partial_i R
+\tfrac16\lsp F^{\smash{\phi}}_{ab}\lsp
\partial^i\phi^a\lsp\partial_i\phi^b\lsp R
+\tfrac12\lsp G^{\smash{\phi}}_{ab}\lsp
\partial_i\phi^a\lsp\partial_j\phi^b\lsp G^{ij}\\
&\hspace{1cm}+\tfrac12\lsp A^{\smash{\phi}}_{ab}\lsp\nabla^2\phi^a\lsp
\nabla^2\phi^b
+\tfrac12\lsp B_{abc}^{\smash{\phi}}\lsp\partial^i\phi^a\lsp\partial_i
\phi^b\lsp\nabla^2\phi^c+\tfrac14\lsp C_{abcd}^\phi\lsp\partial^i\phi^a\lsp
\partial_i\phi^b\lsp\partial^j\phi^c\lsp\partial_j\phi^d\big)\\
&\quad-\smash{\int}\dfour y\sqrt{\hm}\,\partial_i\sigma\lsp\big(
W^{\smash{\phi}}_a\lsp \partial_j\phi^a\lsp G^{ij}
+\tfrac13\lsp\partial^i(DR)
+\tfrac13\lsp Y^{\smash{\phi}}_a\lsp\partial^i\phi\lsp R\\
&\hspace{1cm}+\partial^i(U^{\smash{\phi}}_a\lsp\nabla^2\phi^a
+\tfrac12\lsp V^{\smash{\phi}}_{ab}\lsp\partial^j\phi^a\lsp
\partial_j\phi^b)
+S^{\smash{\phi}}_{ab}\lsp\partial^i\phi^a\lsp\nabla^2\phi^b
+\tfrac12\lsp T^{\smash{\phi}}_{abc}\lsp \partial^j\phi^a\lsp
\partial_j\phi^b\lsp\partial^i\phi^c\big)\,,}[OsbEq]
and it is now straightforward to match coefficients between \HolOsbEq and
\OsbEq.\foot{The results can be found in Appendix \ref{appCCs}.} For
example, we find
\eqn{A=C=-\frac{1}{2\lsp\tilde{\kappa}_{5}^2}\frac{3}{W}\lsp\Phi^2\,.}[]
The fact that $A=C$ is a consequence of using only Einstein gravity in the
bulk. Since $W<0$ we have $A,C>0$. We also have
\eqn{W^{\smash{\phi}}_a=\frac{1}{2\lsp\tilde{\kappa}_{5}^2}\frac{12}{W}
\lsp\Phi\lsp\partial_a\Phi\,,\qquad
G^{\smash{\phi}}_{ab}=-\frac{1}{2\lsp\tilde{\kappa}_{5}^2}
\frac{12}{W}\big((M_{ab}+\partial_a\log W\lsp\partial_b\Phi
+\partial_b\log W\lsp\partial_a\Phi)\lsp\Phi
-\partial_a\Phi\lsp\partial_b\Phi\big)\,,}[WG]
and we find that the consistency condition
\eqn{\partial_a \tilde{A}=\tfrac18\lsp(G^{\smash{\phi}}_{ab}
+\partial_a W^{\smash{\phi}}_b-\partial_b W^{\smash{\phi}}_a)\lsp\beta^b
\,,\qquad \tilde{A}=A+\tfrac18\lsp W^{\smash{\phi}}_a\beta^a\,,}[]
is satisfied with the use of \wZeroTwoEqII and \wZeroTwoEqIV.\foot{The
remaining consistency conditions of \cite{Osborn:1991gm} are also
satisfied---see Appendix \ref{appCCs}.} We also find that when the
$\phi$-dependence of the various quantities is neglected, then
$G^{\smash{\phi}}_{ab}$ is positive-definite due to \wZeroTwoEqII and
\wZeroTwoEqIII if we take $W<0$ and $H_{ab}$ to be positive-definite. This
gives results discussed in~\cite{Anselmi:2000fu, Erdmenger:2001ja},
although the connection to these papers if the $\phi$-dependence is
maintained is not clear.  Positivity of $G^{\smash{\phi}}_{ab}$ in
perturbative field theory has been established in \cite{Jack:1990eb}.

We also have
\eqn{A^{\smash{\phi}}_{ab}=\frac{1}{2\lsp\tilde{\kappa}_{5}^2}
\frac{6}{W}\lsp M_{ac}M_{bd}\lsp H^{cd}\,,}[]
which is negative-definite as expected from the field-theoretic
analysis~\cite{Jack:1990eb}.  If the $\phi$-dependence in \WG is neglected
then we see using \WPhiM that $G^{\smash{\phi}}_{ab}=-2\lsp
A^{\smash{\phi}}_{ab}$, a relation valid in conformal perturbation theory
in field theory~\cite{Jack:1990eb}.

In $d=6$ one should be able to repeat the analysis above and check
holographically the consistency conditions of \cite{Grinstein:2013cka}.
Here we make a comment related to the metric analogous to
$G^{\smash{\phi}}_{ab}$ in $d=6$, $G^{\smash{\phi}}_{6\lsp ab}$.
In~\cite{Grinstein:2014xba} it was shown in multiflavor $\phi^3$ theory in
$d=6$ that this metric is perturbatively negative-definite around the
trivial fixed point. Furthermore, in~\cite{Osborn:2015rna} it was pointed
out that this metric is proportional to the coefficient of the contribution
$W^{iklm}W^{j}{\!}_{klm}\lsp\partial_i\phi^a\lsp \partial_j\phi^b$ in the
notation of \SixdTA. Using our result \SixdTA we find
\eqn{G^{\smash{\phi}}_{6\lsp ab}=-\frac{1}{160\sqrt{10}}
\frac{L^5}{2\lsp \tilde{\kappa}_7^2}\big(5+\sqrt{25-60\lsp z}\big)^{3/2}
\lsp \tilde{c}\lsp\delta_{ab}\,,}[GSix]
at the fixed point, where $\delta_{ab}$ is the Kronecker delta and we use
the definitions \zabctilde. If $\tilde{c}<0$ this would give us a
positive-definite $G^{\smash{\phi}}_{6\lsp ab}$. However, our bound
\czbound shows that $\tilde{c}$ has an undetermined sign, and so we cannot
determine the sign of $G^{\smash{\phi}}_{6\lsp ab}$ from \GSix at the fixed
point.

Besides the energy-positivity bounds discussed in section \bounds there are
more stringent constraints arising from causality considerations in the
bulk~\cite{Camanho:2014apa}. More specifically, causality violations occur
in the bulk for $c\ne0$, unless there is an infinite tower of massive
higher-spin fields. The significance of this result for the $a$-theorem in
a six-dimensional field theory with nonvanishing $G_{6\lsp
ab}^{\smash{\phi}}$ as in \GSix is unclear.

\subsec{Relevant operators}
In order to include relevant scalar deformations we add scalar fields
$\chi^\alpha$ with nonzero mass $m_{\chi}$. Then, the bulk Lagrangian
becomes
\eqn{\Lag_B^{\gm,\phi,\chi}=V(\phi,\chi)-R+\tfrac{1}{2}\lsp
H_{ab}\lsp\partial_\mu\phi^a\lsp\partial^\mu\phi^b
+H_{\alpha
a}\lsp\partial^\mu\chi^\alpha\lsp\partial_\mu\phi^a+\tfrac{1}{2}\lsp
H_{\alpha\beta}\lsp\partial_\mu\chi^\alpha\lsp\partial^\mu\chi^\beta\,.}[]

For concreteness consider $d=4$ and $\Delta_{\chi}=2$. This corresponds to
operators of dimension two in the dual field theory. Now terms in $\Sloc$
do not have definite weight, but we can still expand
\eqna{W(\phi,\chi)&=W(\phi)+X_\alpha(\phi)\lsp\chi^\alpha
+\tfrac12\lsp U_{\alpha\beta}\lsp\chi^\alpha\chi^\beta+\cdots\,,\\
-\Phi(\phi,\chi)\lsp R&=-\Phi(\phi)R+Y_\alpha(\phi)\lsp\chi^\alpha
R+\cdots\,,\\
\tfrac12\lsp M_{ab}(\phi,\chi)\lsp\partial^i\phi^a
\lsp\partial_i\phi^b&=
\tfrac12\lsp M_{ab}(\phi)\lsp\partial^i\phi^a\lsp\partial_i\phi^b
+\tfrac12\lsp N_{\alpha ab}(\phi)\lsp\chi^\alpha\partial^i\phi^a\lsp
\partial_i\phi^b+\cdots\,.}[]
The action $S_{\text{loc}}$ now contains
\eqn{\Lagloc^{(2)}=-\Phi(\phi)R+X_\alpha(\phi)\lsp\chi^\alpha
+\tfrac{1}{2}\lsp M_{ab}(\phi)\lsp\partial^i\phi^a
\partial_i\phi^b}[Lloctwo]
and
\eqn{\Lagloc^{(4)}=\tfrac12\lsp U_{\alpha\beta}(\phi)\lsp\chi^\alpha
\chi^\beta +M_{\alpha a}(\phi)\lsp\partial^i\chi^\alpha\lsp\partial_i
\phi^a+Y_\alpha(\phi)\lsp\chi^\alpha R
+\tfrac12\lsp N_{\alpha ab}(\phi)\lsp\chi^\alpha\partial^i
\phi^a\lsp\partial_i\phi^b\,.}[Llocfour]
We can also break down \wZeroTwoEqs by weight. As an example, writing
$V(\phi,\chi)=V(\phi)+V_\alpha(\phi)\lsp\chi^\alpha+\cdots$, the weight
zero part of \wZeroTwoEqI gives
\eqn{V=\tfrac12\lsp H^{ab}\lsp\partial_a W\lsp\partial_b W
-\frac{d}{4\lsp(d-1)}W^2+H^{\alpha a} X_\alpha\lsp\partial_a W
+\tfrac12\lsp H^{\alpha\beta}X_\alpha X_\beta\,,}[modFlowEqI]
where $V=V(\phi)$ and $W=W(\phi)$.

Note that $\mathscr{L}_{\text{loc}}^{(4)}$ includes more weight-four terms,
e.g.\ $R^2$, but these correspond to ambiguities in the trace anomaly as
has been explained in~\cite{Fukuma:2000bz}. Contrary to this, the terms we
include in \Llocfour do not correspond to ambiguities, although they are of
weight four. This is because the functional derivative with respect to
$\chi$ that appears now in the flow equation \FlowEqII reduces the weight
by two. This, then, modifies equations \wZeroTwoEqs, and allows us to
determine relations involving the coefficients in \Lloctwo and \Llocfour at
weight zero and two respectively. Equation \modFlowEqI is the result at
weight zero.

With \Lloctwo and \Llocfour it is straightforward to work out the
local Callan--Symanzik equation, and find holographic counterparts for the
quantities considered in the case of scalar relevant operators of dimension
two by Osborn~\cite{Osborn:1991gm}.  For example, the left-hand side of
\wdEq receives new contributions of the form
\eqn{\frac{2\lsp(d-1)}{W}\left(H^{\alpha\beta}\lsp\frac{1}{\sqrt{\hm}}
\frac{\delta S_{\text{loc}}^{(4)}}{\delta\chi^\alpha}
\frac{1}{\sqrt{\hm}}\frac{\delta\Gamma}{\delta\chi^\beta}+
H^{\alpha a}\frac{1}{\sqrt{\hm}}
\frac{\delta S_{\text{loc}}^{(2)}}{\delta\chi^\alpha}
\frac{1}{\sqrt{\hm}}\frac{\delta\Gamma}{\delta\phi^a}\right).}[]
These give rise to $\Deltam_\sigma$ and the shifts in $\DeltaWhat_\sigma$
and $\hat{\Delta}_\sigma^\beta$ in~\cite[Eq.~(3.25)]{Osborn:1991gm}.

If all $\chi$'s have $\Delta_\chi\neq2$, then $S_{\text{loc}}$ would
include
\eqn{\Lagloc^{(4-\Delta_{\chi})}=X_\alpha(\phi)\lsp\chi^\alpha\,,\qquad
\Lagloc^{(8-2\Delta_{\chi})}=\tfrac12\lsp U_{\alpha\beta}(\phi)\lsp
\chi^\alpha\chi^\beta\,.}[]
These will result in extra contributions to the left-hand side of \wdEq of
the form
\eqna{\frac{2\lsp(d-1)}{W}\left(H^{\alpha a}\frac{1}{\sqrt{\hm}}
\frac{\delta S_\text{loc}^{(4-\Delta_\chi)}}{\delta\phi^a}+H^{\alpha\beta}
\frac{1}{\sqrt{\hm}}\frac{\delta S_{\text{loc}}^{(8-2\Delta_{\chi})}}
{\delta\chi^\beta}\right)\frac{1}{\sqrt{\hm}}
\frac{\delta\Gamma}{\delta\chi^\alpha}&=\\
&\hspace{-3cm}\frac{2\lsp(d-1)}{W}\big(H^{\alpha a}\lsp\partial_aX_\beta
+H^{\alpha\gamma}\lsp U_{\gamma\beta}\big)\lsp\chi^\beta
\frac{1}{\sqrt{\hm}}\frac{\delta\Gamma}{\delta\chi^\alpha}\,.}[]
The quantity $\frac{2\lsp(d-1)}{W}\big(H^{\alpha a}\lsp\partial_aX_\beta
+H^{\alpha\gamma}\lsp U_{\gamma\beta}\big)$ gives a holographic derivation
of the operator $D^\alpha{\!}_\beta$ of~\cite[Eq.~(2.12)]{Baume:2014rla}.

Finally, we note here that dimension three vector operators in the boundary
theory can also be added by considering gauge fields in the bulk. A
discussion of this can be found in~\cite{Nakayama:2013ssa}.

\ack{We would like to thank Hong Liu for many valuable discussions and
suggestions. AS would also like to thank Hugh Osborn and Guilherme Pimentel
for insightful comments when he presented part of this work at DAMTP,
University of Cambridge, and Kostas Skenderis for helpful discussions.  For
our computations we have relied on \emph{Mathematica} and the package
\href{http://www.xact.es}{\texttt{xAct}}. AS is grateful to the Aspen
Center for Physics (partially supported by National Science Foundation
Grant No.~1066293), for hospitality during the final stages of this
project. SR is supported by the U.S.\ Department of Energy under
cooperative research agreement Contract  No.\ DE-FG02-05ER41360. The
research of AS is supported in part by the National Science Foundation
under Grant No.~1350180. YZ is supported by ARO under Grant No.\
W911NF-12-0486.}


\begin{appendices}
\newsec{ADM formalism}[appADM]
We consider a $(d+1)$-dimensional manifold with coordinates $x^\mu$ and
metric $\gm_{\mu\nu}$. We will work in Euclidean signature so that locally
$\gm_{\mu\nu}=\delta_{\mu\nu}$, where $\delta_{\mu\nu}$ is the Kronecker
delta. The line element can be written in the ADM form
\cite{Arnowitt:1962hi}
\eqn{ds^2=\gm_{\mu\nu}\lsp dx^\mu dx^\nu
=N^2(y,r)\lsp dr^2 +\hm_{ij}\big(dy^i+N^i(y,r)\lsp dr\big)
\big(dy^j+N^j(y,r)\lsp dr\big)\,,}[metADM]
where we assume a hypersurface-foliation of the $(d+1)$-dimensional
spacetime along the radial coordinate $r$, and we define
\eqn{\hm_{ij}=\gm_{\mu\nu}\lsp e_i^\mu e_j^\nu\,,\qquad
e_i^\mu=\frac{\partial x^\mu}{\partial y^i}\,,}[]
as the induced metric and $y$ as the coordinates on the hypersurfaces,
while $N$ is the lapse function and $N^i$ the shift vector.  We also define
the vector
\eqn{t^\mu=\frac{\partial x^\mu}{\partial r}=N n^\mu+N^i e_i^\mu\,,
}[tdefn]
where $n^\mu$ is a vector normal to the hypersurfaces with $n^\mu n_\mu=1$.
The inverse of $\gm_{\mu\nu}$ is given by
\eqn{\gm^{\mu\nu}=\hm^{ij}e_i^\mu e_j^\nu+n^\mu n^\nu\,,}[invg]
where $\hm^{ij}$ is the inverse of $\hm_{ij}$.

The starting point for the decomposition of curvature tensors is the
Gauss--Weingarten equation,
\eqn{\nabla_\nu e_i^\mu \lsp e^\nu_j=\Gamma^{k}{\!}_{ij}\lsp
e_k^\mu-K_{ij}\lsp n^\mu\,,}[GWeq]
as well as the equations
\twoseqn{\nabla^\mu n_\nu\lsp e_i^\nu&=K_{i}{\!}^j e_j^\mu+a_i
n^\mu\,,}[GWlikeforna]
{\nabla_\nu n^\mu\lsp e^\nu_i&=K_{i}{\!}^j e_j^\mu\,,}[GWlikefornb][GWlikeforn]
where
\eqn{K_{ij}=\nabla_{(\mu} n_{\nu)}\lsp e_i^\mu
e_j^\nu=\tfrac12\lsp\Lien\gm_{\mu\nu}\lsp e_i^\mu e_j^\nu\,,}[extcurv]
with $\Lien$ the Lie derivative along $n^\mu$, is the extrinsic curvature,
$\Gamma^i{\!}_{jk}$ is the Christoffel symbol defined from $\hm_{ij}$, and
$a_i=a_\mu e^\mu_i$ with $a_\mu=\nabla_\nu n_\mu\lsp n^\nu$.  With the help
of \GWeq and \GWlikeforn, and with the definition
$[\nabla_\mu,\nabla_\nu]A^\rho=R^{\rho}{\!}_{\sigma
\hspace{-0.7pt}\mu\nu}A^\sigma$ for the Riemann tensor, we can derive
\threeseqn{R_{\mu\nus\rho\sigma} e_i^\mu e_j^\nu e_k^\rho e_l^\sigma&=
R_{ijkl} - K_{ik}K_{jl} + K_{il}K_{jk}\,,}[Riemeeee]
{R_{\mu\nus\rho\sigma}n^\mu e_i^\nu e_j^\rho e_k^\sigma &=
\nabla_k K_{ij} -\nabla_jK_{ik}\,,}[Riemneee]
{R_{\mu\nus\rho\sigma}n^\mu e_i^\nu n^\rho e_j^\sigma &=
-\Lien K_{ij}+K_{ik} K^k{\!}_j+\nabla_i a_j-a_ia_j\,,}[Riemnene][RiemDecomp]
where $\Lien K_{ij}\equiv\Lien K_{\mu\nu}\, e_i^\mu e_j^\nu$,
$K_{\mu\nu}=\nabla_\mu n_\nu-n_\mu a_\nu$, and
$\nabla_i a_j\equiv\nabla_\mu a_\nu \lsp e_i^\mu e_j^\nu$.  Equations
\Riemeeee and \Riemneee are known as the Gauss--Codazzi equations, while
equation \Riemnene is known as the Ricci equation. The decomposition of the
Ricci tensor $R_{\mu\nu}=R^\rho{\!}_{\mu\rho\nu}$ is given by
\threeseqn{R_{\mu\nu} e_i^\mu e_j^\nu&=R_{ij}-\Lien K_{ij}
-KK_{ij}+2\lsp K_{i}{\!}^kK_{kj} +\nabla_i a_j-a_ia_j\,,}[Ricciee]
{R_{\mu\nu}n^\mu e_i^\nu&=-\partial_i K+\nabla_j K^j{\!}_i\,,}[Riccine]
{R_{\mu\nu}n^\mu n^\nu&=-\hm^{ij}\Lien K_{ij}+K^{ij}K_{ij}+\nabla^i a_i-a^i
a_i\,.}[Riccinn][RicciDecomp]
Finally, the Ricci scalar $R=\gm^{\mu\nu}R_{\mu\nu}$ has the
decomposition
\eqn{R={}^d\!R -2\lsp \hm^{ij}\Lien K_{ij} -K^2+3\lsp K^{ij}K_{ij}
+2\lsp(\nabla^i a_i-a^i a_i)\,.}[RicciScDecomp]

In terms of the lapse it is not hard to see that $a_i=-\partial_i\log N$,
and so
\eqn{\nabla_i a_j-a_ia_j=-\frac{1}{N}\nabla_i\partial_jN\,.}[]
It is now possible to express some of the above quantities using the Lie
derivative of the vector $t^\mu$ of \tdefn as opposed to $n^\mu$, as well
as the lapse and shift, since $\Liet=N\Lien+\LieN$.  From \tdefn we see
that $\Liet=\partial/\partial r$ on a scalar. We define the tensor
\eqn{L_{ij}=\Lien K_{ij}+\frac{1}{N}\nabla_i\partial_j N=\frac{1}{N}
\big(\Kdot_{ij}-\LieN K_{ij}+\nabla_i\partial_j N\big)\,,\qquad
\Kdot_{ij}\equiv\Liet K_{ij}\,,}[Ldefn]
with the aid of which we can express \Riemnene as
\eqn{\tag{\ref{Riemnene}$'$}
R_{\mu\nus\rho\sigma}n^\mu e_i^\nu n^\rho e_j^\sigma = K_{i}{\!}^k
K_{kj}-L_{ij}\,,}[Riemneneprime]
and also \Ricciee, \Riccinn and \RicciScDecomp as
\eqn{\tag{\ref{Ricciee}$'$}
R_{\mu\nu} e_i^\mu e_j^\nu=R_{ij}-KK_{ij}+2\lsp K_i{\!}^k K_{kj}
-L_{ij}\,,}[Riccineprime]
\eqn{\tag{\ref{Riccinn}$'$}
R_{\mu\nu}n^\mu n^\nu=K^{ij}K_{ij}-L\,,\qquad L=\hm^{ij}L_{ij}\,,
}[Riccinnprime]
and
\eqn{\tag{\ref{RicciScDecomp}$'$}
R={}^d\! R-K^2+3\lsp K^{ij}K_{ij}-2\lsp L\,.}[RicciScDecompP]

Now, from \extcurv we find
\eqn{\hmdot_{ij}\equiv\Liet\hm_{ij}=2\lsp NK_{ij}+\nabla_i N_j+
\nabla_j N_i\,,}[Liethm]
and we can also compute
\eqn{\Liet\sqrt{\hm}=\sqrt{\hm}\lsp(NK+\nabla^i N_i)\,.}[Lietdethm]
Finally,
\eqn{L=\frac{1}{N}\left(\frac{1}{\sqrt{\hm}}\Liet
\big(\hspace{-0.5pt}\sqrt{\hm}\lsp K\big)
+\nabla^i(\partial_i N-KN_i)\right)-K^2+2\lsp K^{ij}K_{ij}\,,}[]
and so we can write
\eqn{\tag{\ref{RicciScDecomp}$''$}
R={}^d\!R+K^2-K^{ij}K_{ij}-\frac{2}{N}\left(\frac{1}{\sqrt{\hm}}
\Liet\big(\hspace{-0.5pt}\sqrt{\hm}\lsp K\big)+\nabla^i(\partial_i N
-KN_i)\right).}[RicciScDecompDP]
With \RicciScDecompDP we can easily read off the Gibbons--Hawking--York
term for Einstein gravity \cite{Gibbons:1976ue, York:1972sj}.

\newsec{Boundary terms}[appBound]
In this appendix we work out the allowed $\phi$-dependent boundary terms
$\Lag_\partial^\phi$ in \BoundaryTerms. The form of $\Lag_\partial^\hm$ has
been determined in \cite{Fukuma:2001uf}.

The transformation properties of a general symmetric two-index tensor
$S_{\mu\nu}$ with the ADM decomposition
\eqn{S_{\mu\nu}\lsp dx^\mu dx^\nu = F(y,r) \lsp dr^2
+ 2\lsp G_i(y,r)\lsp dr\lsp dy^i+H_{ij}(y,r)\lsp dy^i\lsp dy^j}[]
under the infinitesimal transformation considered in
\cite{Fukuma:2002sb}, namely
\eqna{r&\rightarrow r' = r+\epsilon(y,r)\,,\\
y^i&\rightarrow y^{\prime\lsp i}= y^i+\epsilon^i(y,r)\,,}[infdiffeo]
can be seen as follows.\foot{Note that under a finite diffeomorphism of the
form $x^\mu\to f^\mu(x)$ we require $f^r(y,r_0)=r_0$ so that the location
of the boundary is fixed.}  Under a change of coordinates and demanding
that $S_{\mu\nu}$ be invariant we get the condition
\eqn{F\lsp dr^2 + 2\lsp G_i\lsp dr\lsp dy^i+H_{ij}\lsp dy^i\lsp dy^j =
F'\lsp dr^{\prime\lsp 2} + 2\lsp G'_i\lsp dr' dy^{\prime\lsp i}
+H'_{ij}\lsp dy^{\prime\lsp i}\lsp dy^{\prime\lsp j}\,,}[]
where the primed quantities on the right-hand side are functions of
$r',y'$.  Also,
\eqna{ dr' &= dr +\partial_r \epsilon\lsp dr
+ \partial_i \epsilon\lsp dy^i\,,\\
dy^{\prime\lsp i} &= dy^i + \partial_r \epsilon^i dr + \partial_j
\epsilon^i\lsp dy^j\,,}[]
and from the first equation we see that when $dr=0$ then $dr'=0$ only if
$\partial_i\epsilon=0$. This is a condition we require only on the
boundary, and thus $\partial_i\epsilon(y,r_0)=0$.  Keeping only terms
involving the derivatives of $\epsilon,\epsilon^i$, we get
\eqna{
F' &= F-2\lsp F\lsp\partial_r\epsilon-2\lsp G_i\lsp \partial_r\epsilon^i\,,
\\
G'_i &= G_i - F\lsp\partial_i\epsilon-G_i\lsp\partial_r\epsilon
-G_j\lsp\partial_i\lsp\epsilon^j-H_{ij}\lsp\partial_r\epsilon^j\,,\\
H'_{ij} &= H_{ij}-G_i\lsp\partial_j\epsilon-G_j\lsp\partial_i\epsilon
-H_{ik}\lsp\partial_j\epsilon^k-H_{jk}\lsp\partial_i\epsilon^k\,.}[]
As a check, we note that taking $H_{ij} = \hm_{ij}$, $F =
N^2+\hm_{ij}N^iN^j$ and $G_i = N_i$ reproduces equation (C.2) of
\cite{Fukuma:2002sb} for the transformation of the lapse, shift and induced
metric.

We now apply this procedure to the tensor $\partial_\mu\phi\lsp
\partial_\nu\phi$, which has the ADM form
\eqn{\partial_\mu\phi\lsp\partial_\nu\phi = (N\Lien\phi+\LieN\phi)^2
N^2\lsp dr^2 +
2\lsp(N\Lien\phi+\LieN\phi)\lsp\lsp\partial_i\phi\lsp\lsp dr\lsp dy^i
+\partial_i\phi\lsp\partial_j\phi\, dy^i\lsp dy^j\,,}[]
and get
\eqn{\delta(\partial_i\phi\lsp\partial_j\phi) =
-\partial_i\phi\lsp\partial_k\phi\lsp\partial_j\epsilon^k
-\partial_j\phi\lsp\partial_k\phi\lsp\partial_i\epsilon^k\,.}[varDelPhisq]
Also, if we use $\partial_i \epsilon(y,r_0) = 0$ then
\eqn{\delta K_{ij} = -\partial_i\epsilon^k K_{kj}-\partial_j\epsilon^k
K_{ki}\,,}[]
and so
\eqn{\delta K^{ij} = -(\hm^{ik}K^{jl}+g^{jk}K^{il})\delta \hm_{kl} +
g^{ik}g^{jl}\lsp \delta K_{jl} = K^{ik}\partial_k\lsp\epsilon^j +
K^{jk}\lsp\partial_k\epsilon^i\,.}[varKinv]
Thus, with the use of \varDelPhisq and \varKinv we finally find
\eqna{\delta(K^{ij}\partial_i\phi\lsp\partial_j\phi) = 0\,,}[]
which shows that $K^{ij}\partial_i\phi\lsp\partial_j\phi$ is an allowed
boundary term. Furthermore, since $\delta K = 0$ and $
\delta(\partial^i\phi\lsp\partial_i\phi)=0$, we conclude that
$K\lsp\partial^i\phi\lsp\partial_i\phi$ is also allowed.

There is a further term that is quadratic in $\phi$, has three derivatives
at the boundary and one of them is a radial derivative, namely
$\partial^i\!\Lien\phi\lsp\lsp\partial_i\phi$, or, equivalently,
$\Lien\phi\lsp\nabla^2\phi$. The transformation property of the Lie
derivative can be worked out by considering
\eqn{d\phi=(N\Lien\phi+\LieN\phi)\lsp\lsp dr + \partial_i\phi\lsp\lsp dy^i=
(N\Lien\phi+\LieN\phi)' dr' + (\partial_i\phi)'\lsp dy^{\prime\lsp i}\,,}[]
which gives
\eqn{\delta(N\Lien\phi) =
-N\Lien\phi\lsp\partial_r\epsilon
\Rightarrow\delta(\Lien\phi) = 0\,.}[varLieD]
Furthermore, we have $\delta(\nabla_i\partial_j\phi) =
-\nabla_i\partial_k\phi\lsp\partial_j\epsilon^k
-\nabla_j\partial_k\phi\lsp\partial_i\epsilon^k$ and so
\eqna{\delta(\nabla^2\phi) = 0\,.}[varNabla]
Thus, from \varLieD and \varNabla we see that the term
$\Lien\phi\lsp\nabla^2\phi$ is an allowed boundary term. Note that the
difference between the terms $\nabla^2\phi\lsp\nabla^2\phi$ and
$\nabla^\mu\partial^\nu\phi\lsp \nabla_\mu\partial_\nu\phi$ in \BulkTerms
contains just the boundary terms in $\mathscr{L}_\partial^\phi$ of
\BoundaryTerms. Nevertheless, the coefficient of
$\nabla^\mu\partial^\nu\phi\lsp \nabla_\mu\partial_\nu\phi$ in \BulkTerms
contributes to the anomaly \SixdTA while that of
$\nabla^2\phi\lsp\nabla^2\phi$ does not.

\newsec{Expressions for six-dimensional curvature tensors}[appSixdCurv]
A complete basis of scalar dimension-six curvature terms consists of \cite{Bonora:1985cq}
\eqn{\begin{gathered}
K_1=R^3\,,\qquad
K_2=RR^{ij}R_{ij}\,,\qquad
K_3=RR^{ijkl}R_{ijkl}\,,\qquad
K_4=R^{ij}R_{jk}R^k_{\hphantom{k}\!i}\,,\\
K_5=R^{ij}R^{kl}R_{iklj}\,,\qquad
K_6=R^{ij}R_{iklm}
  R_{j}^{\smash{\hphantom{j}klm}}\,,\qquad
K_7=R^{ijkl}R_{klmn}
R^{mn}{\!}_{ij}\,,\\
K_8=R^{ijkl}R_{mjkn}R_{i}{}^{mn}{}_l\,,\qquad
K_9=R\,\nabla^2 R\,,\qquad
K_{10}=R^{ij}\,\nabla^2 R_{ij}\,,\qquad
K_{11}=R^{ijkl}\,\nabla^2 R_{ijkl}\,,\\
K_{12}=R^{ij}\nabla_i\partial_j R\,,\qquad
K_{13}=\nabla^i R^{jk}\lsp\nabla_{i}R_{jk}\,,\qquad
K_{14}=\nabla^i R^{jk}\lsp\nabla_{j}R_{ik}\,,\\
K_{15}=\nabla^i R^{jklm} \lsp
\nabla_i R_{jklm}\,,\qquad
K_{16}=\nabla^2 R^2\,,\qquad
K_{17}=(\nabla^2)^2 R\,.
\end{gathered}}
In equation \SixdTA in the main text we use the combinations
\eqna{I_1&=\tfrac{19}{800}\lsp K_1 - \tfrac{57}{160}\lsp K_2 + \tfrac{3}{40}\lsp K_3 +
\tfrac{7}{16}\lsp K_4 - \tfrac{9}{8}\lsp K_5 - \tfrac{3}{4}\lsp K_6 +
\lsp K_8\,,\displaybreak[0]\\
I_2&=\tfrac{9}{200}\lsp K_1 - \tfrac{27}{40}\lsp K_2 + \tfrac{3}{10}\lsp K_3 +
\tfrac{5}{4}\lsp K_4 - \tfrac{3}{2}\lsp K_5 - 3\lsp K_6 + \lsp K_7\,,\displaybreak[0]\\
I_3&=-\tfrac{11}{50}\lsp K_1 + \tfrac{27}{10}\lsp K_2 - \tfrac{6}{5}\lsp K_3 - \lsp K_4 + 6\lsp K_5+
2\lsp K_7 - 8\lsp K_8\\&\hspace{3cm} + \tfrac{3}{5}\lsp K_9 - 6\lsp K_{10} + 6\lsp K_{11} + 3\lsp K_{13}
- 6\lsp K_{14} + 3\lsp K_{15}\,,\displaybreak[0]\\
E_6&=\lsp K_1 - 12\lsp K_2 + 3\lsp K_3 + 16\lsp K_4 - 24\lsp K_5 - 24\lsp K_6 + 4\lsp K_7 +
8\lsp K_8\,,\displaybreak[0]\\
J_1&=6\lsp K_6 - 3\lsp K_7 + 12\lsp K_8 + \lsp K_{10} - 7\lsp K_{11} - 11\lsp K_{13} + 12\lsp K_{14} -
  4\lsp K_{15}\,,\displaybreak[0]\\
J_2&=-\tfrac15 \lsp K_9 + \lsp K_{10} + \tfrac25 \lsp K_{12} + \lsp
K_{13}\,,\displaybreak[0]\\
J_3&=\lsp K_4 + \lsp K_5 - \tfrac{3}{20}\lsp K_9 + \tfrac45 \lsp K_{12}
  + \lsp K_{14}\,,\displaybreak[0]\\
J_4&=-\tfrac15 \lsp K_9 + \lsp K_{11} + \tfrac25 \lsp K_{12} + \lsp
K_{15}\,.}

\newsec{Anomaly coefficients and consistency conditions in
\texorpdfstring{$d=4$}{d=4}}[appCCs]
Osborn's consistency conditions in $d=4$ can all be verified using the
holographic result \HolOsbEq. First, we list here the holographic results
for the coefficients in \OsbEq. We have
\eqn{A=C=-\frac{1}{2\lsp\tilde{\kappa}_{5}^2}\frac{3}{W}\lsp\Phi^2,\qquad
B=\frac{1}{2\lsp\tilde{\kappa}_{5}^2}\frac{27}{W}\lsp\partial_a\Phi\lsp
\partial_b\Phi\lsp H^{ab}\,,\qquad D=0\,,}[]
\eqn{E^{\smash{\phi}}_a=-\frac{1}{2\lsp\tilde{\kappa}_5^2}\lsp
\frac{18}{W}\lsp M_{ab}\lsp\partial_c\Phi\lsp H^{cd}\,,\quad
W^{\smash{\phi}}_a=\frac{1}{2\lsp\tilde{\kappa}_5^2}\lsp\frac{12}{W}\lsp
\Phi\lsp\partial_a\Phi\,,\quad
Y^{\smash{\phi}}_a=-\frac{1}{2\lsp\tilde{\kappa}_5^2}\lsp\frac{18}{W}\lsp
M_{ab}\lsp\partial_c\Phi\lsp H^{bc}\,,\quad
U^{\smash{\phi}}_a=0\,,}[]
\eqna{F^{\smash{\phi}}_{ab}&=-\frac{1}{2\lsp\tilde{\kappa}_5^2}\lsp
\frac{6}{W}\big(M_{ab}\Phi+3\lsp\partial_a\Phi\lsp\partial_b\Phi
+3\lsp\partial_c M_{ab}\lsp\partial_d\Phi\lsp H^{cd}
-6\lsp M_{c(a}\lsp\partial_{b)}\log W\lsp\partial_d\Phi\lsp H^{cd}\\
&\hspace{7cm}+6\lsp M_{c(a}\lsp\partial_{b)}\partial_d\Phi\lsp H^{cd}
+6\lsp M_{c(a}\lsp\partial_{b)}H^{cd}\lsp\partial_d\Phi\big)\,,\\
G^{\smash{\phi}}_{ab}&=-\frac{1}{2\lsp\tilde{\kappa}_{5}^2}
\frac{12}{W}\big((M_{ab}+\partial_a\log W\lsp\partial_b\Phi
+\partial_b\log W\lsp\partial_a\Phi)\lsp\Phi
-\partial_a\Phi\lsp\partial_b\Phi\big)\,,\\
A^{\smash{\phi}}_{ab}&=\frac{1}{2\lsp\tilde{\kappa}_{5}^2}
\frac{6}{W}\lsp M_{ac}M_{bd}\lsp H^{cd}\,,\qquad
V^{\smash{\phi}}_{ab}=-\frac{1}{2\lsp\tilde{\kappa}_{5}^2}
\frac{6}{W}\lsp \partial_a\Phi\lsp\partial_b\Phi\,,\qquad
S^{\smash{\phi}}_{ab}=\frac{1}{2\lsp\tilde{\kappa}_{5}^2}
\frac{6}{W}\lsp\partial_a\Phi\lsp\partial_b\Phi\,,}[]
\eqna{B^{\smash{\phi}}_{abc}&=-\frac{1}{2\lsp\tilde{\kappa}_5^2}
\frac{6}{W}\big(\partial_a\Phi\lsp\partial_b\Phi\lsp\partial_c\log W
+2\lsp \partial_c\Phi\lsp\partial_{(a}\Phi\lsp\partial_{b)}\log W
+2\lsp M_{c(a}\lsp\partial_{b)}\Phi-2\lsp \Gamma^M_{abd}\lsp M_{ce}\lsp
H^{de}\big)\,,\\
T^{\smash{\phi}}_{abc}&=\frac{1}{2\lsp\tilde{\kappa}_5^2}\frac{6}{W}
\big(M_{ab}\lsp\partial_c\Phi
-2\lsp\partial_a\Phi\lsp\partial_b\Phi\lsp\partial_c\log W
+2\lsp\partial_a\partial_b\Phi\lsp\partial_c\Phi
-2\lsp M_{c(a}\lsp\partial_{b)}\Phi\big)\,,}[]
and
\eqna{C^{\smash{\phi}}_{abcd}=-\smash{\frac{1}{2\lsp\tilde{\kappa}_5^2}
\frac{2}{W}}\big(&M_{ab}\lsp M_{cd}-3\lsp M_{a(c}\lsp M_{d)b}\\
&+3\lsp M_{ab}\lsp\partial_{(c}\Phi\lsp\partial_{d)}\log W
+3\lsp M_{cd}\lsp\partial_{(a}\Phi\lsp\partial_{b)}\log W\\
&-3\lsp M_{ac}\lsp\partial_{(b}\Phi\lsp\partial_{d)}\log W
-3\lsp M_{bc}\lsp\partial_{(a}\Phi\lsp\partial_{d)}\log W\\
&-3\lsp M_{ad}\lsp\partial_{(b}\Phi\lsp\partial_{c)}\log W
-3\lsp M_{bd}\lsp\partial_{(a}\Phi\lsp\partial_{c)}\log W\\
&-3\lsp\partial_a\Phi\lsp\partial_b\Phi\lsp\partial_c\log W\lsp
\partial_d\log W
-3\lsp\partial_c\Phi\lsp\partial_d\Phi\lsp\partial_a\log W\lsp
\partial_b\log W\\
&+6\lsp\partial_a\partial_b\Phi\lsp\partial_{(c}\Phi\lsp\partial_{d)}\log W
+6\lsp\partial_c\partial_d\Phi\lsp\partial_{(a}\Phi\lsp\partial_{b)}\log
W\\
&+3\lsp\partial_a\Phi\lsp\partial_b\Phi\lsp\partial_c\partial_d\log W
+3\lsp\partial_c\Phi\lsp\partial_d\Phi\lsp\partial_a\partial_b\log W\\
&+6\lsp\partial_d M_{c(a}\lsp\partial_{b)}\Phi
+6\lsp\Gamma^M_{ab(c}\lsp\partial_{d)}\Phi
-6\lsp\partial_{(a}\Phi\lsp\Gamma^M_{b)cd}
-6\lsp \Gamma^M_{abe}\lsp\Gamma^M_{cdf}\lsp H^{ef}\big)\,.}[]
With these results and with the use of \wZeroTwoEqII, \wZeroTwoEqIII,
\wZeroTwoEqIV, and \holbeta we find that the consistency conditions of
\cite{Osborn:1991gm}, namely
\eqna{8\lsp\partial_a A-G^{\smash{\phi}}_{ab}\lsp\beta^b&=
-\Liebeta W^{\smash{\phi}}_a\,,\\
2\lsp E^{\smash{\phi}}_i+A^{\smash{\phi}}_{ab}\lsp\beta^b&=
-\Liebeta U^{\smash{\phi}}_a\,,\\
8\lsp B-A^{\smash{\phi}}_{ab}\lsp\beta^a\beta^b&=
\Liebeta(2\lsp D+U^{\smash{\phi}}_a\beta^a)\,,\\
4\lsp\partial_a B+(A^{\smash{\phi}}_{ab} + F^{\smash{\phi}}_{ab})\beta^b
&=\Liebeta(\partial_a D+Y^{\smash{\phi}}_a-U^{\smash{\phi}}_a)\,,\\
G^{\smash{\phi}}_{ab}+2\lsp A^{\smash{\phi}}_{ab}
+\Lambda^{\smash{\phi}}_{ab}&=\Liebeta S^{\smash{\phi}}_{ab},\qquad
\Lambda^{\smash{\phi}}_{ab}=2\lsp\partial_a\beta^c A^{\smash{\phi}}_{cb}+
\beta^c B^{\smash{\phi}}_{cab}\,,\\
2(A^{\smash{\phi}}_{ab}+F^{\smash{\phi}}_{ab})+\Lambda^{\smash{\phi}}_{ab}
+\beta^c(2\lsp\bar{A}^{\smash{\phi}}_{c(ab)}-\bar{A}^{\smash{\phi}}_{abc})
&=\Liebeta(S^{\smash{\phi}}_{ab}-A^{\smash{\phi}}_{ab}-2\lsp
\partial_{(a}U^{\smash{\phi}}_{b)}+V^{\smash{\phi}}_{ab})\,,\\
&\hspace{3cm}\bar{A}^{\smash{\phi}}_{abc}=\partial_c A^{\smash{\phi}}_{ab}-
B^{\smash{\phi}}_{c(ab)}\,,\\
\partial_{(a}G^{\smash{\phi}}_{b)c}
-\tfrac12\partial_c G^{\smash{\phi}}_{ab}
+B^{\smash{\phi}}_{abc}+\partial_c\beta^d
B^{\smash{\phi}}_{abd} +C^{\smash{\phi}}_{abcd}\lsp\beta^d&=\tfrac12
\Liebeta T^{\smash{\phi}}_{abc}+\partial_a\partial_b\beta^d
S^{\smash{\phi}}_{cd}\,,}[allCCs]
are satisfied.

\end{appendices}

\bibliography{hol_anom_dilaton_gravity}

\end{document}